\documentclass{article}
\usepackage{slashed,verbatim,subfigure}
\usepackage[numbers,sort&compress]{natbib}
\usepackage{amsmath}
\usepackage{amssymb}
\usepackage{appendix}
\usepackage{graphicx}
\usepackage{dcolumn}
\usepackage{bm}
\usepackage[colorlinks=true,linktocpage=true,
linkcolor=blue,citecolor=blue]{hyperref}
\usepackage[a4paper]{geometry}
\newcommand{\be}{\begin{eqnarray}}
\newcommand{\ee}{\end{eqnarray}}
\newcommand{\ec}{\sigma_{\rm el}}
\newcommand{\eh}{\sigma_{\rm H}}
\begin{document}
\large
\title{\bf{Effects of weak magnetic field and finite chemical potential on the transport of charge and heat in hot QCD matter}}
\author{Shubhalaxmi Rath\footnote{shubhalaxmi@iitb.ac.in}~~and~~Sadhana Dash\footnote{sadhana@phy.iitb.ac.in}\vspace{0.03in} \\ Department of Physics, Indian Institute of Technology Bombay, Mumbai 400076, India}
\date{}
\maketitle
\begin{abstract}
We have studied the effects of weak magnetic field and finite chemical potential 
on the transport of charge and heat in hot QCD matter by estimating their 
respective response functions, {\em viz.} the electrical conductivity 
($\sigma_{\rm el}$), the Hall conductivity ($\sigma_{\rm H}$), the thermal 
conductivity ($\kappa_0$) and the Hall-type thermal conductivity ($\kappa_1$). 
The expressions of charge and heat transport coefficients are 
obtained by solving the relativistic Boltzmann transport equation 
in the relaxation time approximation at weak magnetic field and finite chemical 
potential. The interactions among partons are incorporated through their 
thermal masses. We have observed that $\ec$ and 
$\kappa_0$ decrease and $\eh$ and $\kappa_1$ increase with the magnetic field 
in the weak magnetic field regime. On the other hand, the presence of a finite chemical 
potential increases these transport coefficients. The effects of weak magnetic field 
and finite chemical potential on aforesaid transport coefficients are found to be 
more conspicuous at low temperatures, whereas at high temperatures, they 
have only a mild dependence on magnetic field and chemical potential. We have 
found that the presence of finite chemical potential further extends the lifetime 
of the magnetic field. Furthermore, we have explored the effects of weak magnetic 
field and finite chemical potential on the Knudsen number, the elliptic flow 
coefficient and the Wiedemann-Franz law. 

\end{abstract}

\newpage

\section{Introduction}
High temperatures and/or high densities create most favorable conditions 
for the transition of the normal nuclear matter to a deconfined state of 
quarks and gluons, known as quark-gluon plasma (QGP). 
Such conditions are realised in collisions of ultrarelativistic heavy ions at 
Relativistic Heavy Ion Collider (RHIC) at BNL and Large Hadron Collider (LHC) at 
CERN. In noncentral collisions, strong magnetic fields are 
produced in a direction perpendicular to the collision plane whose
initial strength could be expressed in terms of the pion mass scale 
as $eB=m_{\pi}^2$ ($\simeq 10^{18}$ Gauss) at RHIC and $eB=15$ $m_{\pi}^2$ at 
LHC \cite{Skokov:IJMPA24'2009,Bzdak:PLB710'2012} energies. Some of the 
phenomenological consequences of the strong magnetic fields are the chiral 
magnetic effect \cite{Fukushima:PRD78'2008,Kharzeev:NPA803'2008}, the axial magnetic 
effect  \cite{Braguta:PRD89'2014,Chernodub:PRB89'2014}, the nonlinear 
electromagnetic current  \cite{Kharzeev:PPNP75'2014,Satow:PRD90'2014}, the 
axial Hall current \cite{Pu:PRD91'2015}, the chiral vortical effect \cite{Kharzeev:PRL106'2011} etc. 
Whether the magnetic field created leaves any observable effects in heavy ion collisions depends on the 
thermalization of light quarks in QGP which induces a large electrical conductivity. As a 
result, an electric current would be induced by virtue of Lenz's law
which might significantly elongate the lifetime of the magnetic field \cite{Tuchin:AHEP2013'2013,McLerran:NPA929'2014,Rath:PRD100'2019}. Therefore, 
it is pertinent to investigate the effects of magnetic field on the partonic 
medium. The observable effects of strong magnetic fields on 
various properties of hot medium of quarks and gluons have attracted much theoretical attention, {\em e.g.} the thermodynamic and magnetic 
properties \cite{Rath:JHEP1712'2017,Bandyopadhyay:PRD100'2019,Rath:EPJA55'2019,
Karmakar:PRD99'2019}, the photon and dilepton productions from QGP \cite{Hees:PRC84'2011,Shen:PRC89'2014,Tuchin:PRC88'2013,Mamo:JHEP1308'2013}, the 
heavy quark diffusion \cite{Fukushima:PRD93'2016}, the magnetohydrodynamics \cite{Roy:PLB750'2015,Inghirami:EPJC76'2016} etc. 

The study of various transport coefficients is important as they 
provide information about the formation and evolution of the 
hot QCD matter. Two of such transport coefficients are the electrical 
conductivity ($\ec$) and the thermal conductivity ($\kappa$) which 
describe the charge transport and the heat transport in the medium, 
respectively. Besides the shear and bulk viscosities, the electrical 
and thermal conductivities are also vital for the 
hydrodynamic evolution of the strongly interacting matter at nonzero 
baryon densities \cite{Kapusta:PRC86'2012,Denicol:PRD89'2014}. These 
conductivities can be determined by using various approaches, {\em viz.}, the 
relativistic Boltzmann transport equation \cite{Muronga:PRC76'2007,Puglisi:PRD90'2014,Thakur:PRD95'2017,
Yasui:PRD96'2017}, the correlator technique using Green-Kubo formula \cite{Nam:PRD86'2012,Greif:PRD90'2014,Feng:PRD96'2017}, the Chapman-Enskog 
approximation \cite{Mitra:PRD94'2016,Mitra:PRD96'2017}, the lattice 
simulation \cite{Gupta:PLB597'2004,Aarts:JHEP1502'2015,Ding:PRD94'2016} etc. 
In the presence of magnetic field, the transport coefficients no longer 
remain isotropic and they acquire multicomponent structures \cite{Lifshitz:BOOK'1981,Harutyunyan:PRC94'2016,Denicol:PRD98'2018,Das:PRD99'2019,
Das:PRD100'2019,Das:PRD101'2020,Chen:PRD101'2020,Dash:PRD102'2020,
Bandyopadhyay:PRD102'2020,Chatterjee:EPJA57'2021,Dey:JP95'2021,
Satapathy:PRD104'2021,Dey:IJMPE30'2021}. There exist three components for 
charge transport as well as for heat transport. However, under 
specific conditions, where electric and magnetic fields are perpendicular to 
each other and in weak magnetic field limit, some components vanish \cite{Feng:PRD96'2017,Das:PRD99'2019,Chatterjee:EPJA57'2021}. 
In the presence of a magnetic field, quarks experience a Lorentz force and it 
results in an induced electric current along a direction transverse to both 
electric and magnetic fields, and the conductivity associated 
with this current is known as Hall conductivity \cite{Feng:PRD96'2017}. 
For pair plasma having equal numbers of charged particles and 
antiparticles, the net Hall current vanishes 
\cite{Blackman:PRL71'1993,Bessho:PP14'2007}. At zero 
chemical potential, quark-gluon plasma is analogous to the case 
of pair plasma. In this case, there will be no Hall effect due 
to the exact cancellation of Hall current contributions from 
particles and their antiparticles. However, at finite chemical 
potential, the asymmetry between the numbers of quarks and 
antiquarks develops a suitable condition for the production of 
finite Hall current. In the weak magnetic field regime, one can 
assume that the phase space and the single particle energies 
are not affected by the magnetic field through Landau quantization 
as in ref. \cite{Feng:PRD96'2017} and the main contribution of the 
effect of magnetic field on the transport coefficients comes through the 
cyclotron frequency. 

The effects of magnetic field on the abovementioned transport coefficients had been studied previously using different models and approximations at finite magnetic field. 
For example, in ref. \cite{Buividovich:PRL105'2010}, the effect of magnetic field on the conductivities had been studied using the quenched SU(2) lattice gauge theory. Authors in references \cite{Das:PRD99'2019,Das:PRD100'2019} had estimated the conductivities using the relativistic Boltzmann transport equation in the relaxation time approximation, but for a hot and dense hadronic matter. In ref. \cite{Nam:PRD86'2012}, authors had exploited the Kubo formalism with the dilute instanton-liquid model to study the electrical conductivity in an external magnetic field. In ref. \cite{Hattori:PRD94'2016}, the real time formalism with the diagrammatic method had been exploited to study the electrical conductivity in strong magnetic fields, whereas ref. \cite{Feng:PRD96'2017} had employed the Kubo formalism and ref. \cite{Thakur:PRD100'2019} had used the quasiparticle model to explore the conductivities. Authors in ref. \cite{Fukushima:PRL120'2018} had investigated the effect of magnetic field on the conductivities using the Landau level resummation via kinetic equations. In ref. \cite{Rath:PRD100'2019}, authors had studied the effects of the strong magnetic field-induced and asymptotic expansion-induced anisotropies on conductivities for a hot QCD matter using the kinetic theory approach, while in ref. \cite{Rath:EPJC80'2020}, the collective effects of the strong magnetic field and density on conductivities had been explored. The effective fugacity approach had been implemented in references \cite{Kurian:PRD99'2019,Kurian:EPJC79'2019} to investigate the effect of magnetic field on conductivities. In the present work, (i) we have studied both charge and heat transport coefficients for a QGP/hot QCD matter in the presence of both weak magnetic field and finite chemical potential. In this study, we have used the weak magnetic field limit, where the energy scale associated with the temperature of the QCD medium is larger than the energy scale related to the magnetic field, {\em i.e.} $T^2 \gg eB$. So, we have used the ansatz method in the weak magnetic field limit to calculate the conductivities in section 2. (ii) We have extended our study to know the collective effects of weak magnetic field and density on some applications of conductivities, such as the Knudsen number, the elliptic flow and the Lorenz number in the Wiedemann-Franz law. (iii) We have used the thermal masses of particles. The particles acquire thermally generated masses due to their interactions with the thermal medium. 

In studying the charge and heat transport coefficients for the hot and dense QCD matter, the kinetic theory approach within the relaxation time approximation has been used. Charge and heat transport coefficients are determined by solving the relativistic Boltzmann transport equation in weak magnetic field 
regime. The impact of weak magnetic field on the properties of QCD medium is expected to be 
significantly different from that of the strong magnetic field case. We also 
observe how the presence of finite chemical potential affects the lifetime of 
magnetic field. The use of thermal masses is relevant, because the medium formed in 
heavy ion collision behaves like a strongly coupled system and thus one cannot fully rely on the perturbative method and the interactions are contained only in the thermal masses of particles. Thermal masses have been calculated previously by different groups for different scenarios, {\em viz.}, the Nambu-Jona-Lasinio (NJL) and Polyakov NJL based quasiparticle models \cite{Fukushima:PLB591'2004,Ghosh:PRD73'2006,Abuki:PLB676'2009}, 
quasiparticle model with Gribov-Zwanziger quantization 
\cite{Su:PRL114'2015,Florkowski:PRC94'2016}, quasiparticle model in a strong 
magnetic field \cite{Rath:PRD102'2020,Rath:EPJC81'2021}, thermodynamically 
consistent quasiparticle model \cite{Bannur:JHEP0709'2007,Bannur:PRC75'2007,Srivastava:PRD82'2010,Srivastava:PRD85'2012} etc. 

Due to the finite electrical conductivity in heavy ion collisions, electric 
current is produced, which is essential for the strength of chiral 
magnetic effect \cite{Fukushima:PRD78'2008}. In addition, the strength 
of the charge asymmetric flow in mass asymmetric collisions depends on 
electrical conductivity \cite{Hirono:PRC90'2014}. The difference 
between energy flow and enthalpy flow in a thermal medium generates 
heat flow, which is associated with the thermal conductivity. In heavy 
ion collisions, thermal conductivity plays an important role in 
controlling the strength of hydrodynamic fluctuations 
\cite{Kapusta:PRC85'2012}. Thus, any modification in the charge and 
heat transport coefficients might leave some noticeable impacts on the 
observables at heavy ion collisions. Some of their applications in the 
similar environment, such as the validity of the local 
equilibrium through the Knudsen number ($\Omega$), the 
elliptic flow ($v_2$) and the interplay between electrical and thermal 
conductivities through the Lorenz number ($L$) in the Wiedemann-Franz law 
have also been explored. 

The rest of this paper is organized as follows. Section 2 is devoted to the study 
of charge and heat transport properties of hot and dense QCD matter in the 
presence of a weak magnetic field by using the kinetic theory approach. The effect 
of chemical potential on the lifetime of magnetic field in an electrically 
conducting medium is discussed in section 3. The results on aforesaid transport 
properties are discussed in section 4. Section 5 explores the applications of the 
obtained transport coefficients to estimate the Knudsen number, the elliptic flow 
coefficient and to study the relation between charge and heat transports through the Wiedemann-Franz law. The work is summarised in section 6. 

\section{Charge and heat transport properties of hot and dense QCD medium in a weak magnetic field}
In this section, the charge and heat transport properties are studied in 
kinetic theory approach by calculating corresponding transport 
coefficients. In particular, subsection 2.1 contains the calculation of 
components of charge transport and subsection 2.2 is devoted to the 
calculation of components of heat transport for the hot and dense QCD 
medium in the presence of a weak magnetic field. 

To calculate the transport coefficients, we solve the relativistic Boltzmann transport equation 
by following the relaxation time approximation. In general, the Boltzmann transport equation is 
a complicated nonlinear integro-differential equation for particle distribution function $f(p)$, 
which gets linearized through the relaxation time approximation, and this can be understood as 
follows. Frequent collisions among particles help the system in bringing back to the equilibrium 
state and in this process the Boltzmann transport equation expresses the evolution of the particle distribution function as
\begin{eqnarray}\label{1}
\frac{df(\mathbf{p})}{dt}=\left(\frac{df(\mathbf{p})}{dt}\right)_{\rm coll}
.\end{eqnarray}
In a time interval $dt$, the probability of occupation of a parton with momentum $\mathbf{p}$ after it gets scattered into the volume element $d\mathbf{p^\prime}$ about $\mathbf{p^\prime}$ is written as $\Gamma_{\mathbf{p},\mathbf{p^\prime}}dtd\mathbf{p^\prime}/(2\pi)^3$, where $\Gamma_{\mathbf{p},\mathbf{p^\prime}}$ is the matrix element discerning the scattering of partons in and out of the aforesaid volume element. Denoting $\tau(\mathbf{p})$ as the relaxation time, the probability per unit time that a parton at $\mathbf{p}$ gets scattered into $\mathbf{p^\prime}$ is given by
\begin{eqnarray}\label{2}
\frac{1}{\tau(\mathbf{p})}=\int\frac{d\mathbf{p^\prime}}{(2\pi)^3}\Gamma_{\mathbf{p},\mathbf{p^\prime}}\left(1-f(\mathbf{p^\prime})\right)
.\end{eqnarray}
The number of partons per unit volume in $d\mathbf{p}$ about $\mathbf{p}$ experiencing a collision 
within a time interval $dt$ is $\frac{dt}{\tau(\mathbf{p})}{f(\mathbf{p})d\mathbf{p}}/{(2\pi)^3}$, 
which can also be represented as ~ $-\left(\frac{df(\mathbf{p})}{dt}\right)_{\rm out}{d\mathbf{p}dt}/{(2\pi)^3}$. Thus, the number of partons going out of the aforesaid volume element after suffering 
collisions is obtained as
\begin{eqnarray}\label{3}
\left(\frac{df(\mathbf{p})}{dt}\right)_{\rm out}=-\frac{f(\mathbf{p})}{\tau(\mathbf{p})}
.\end{eqnarray}
After substituting the value of $1/\tau(\mathbf{p})$ \eqref{2} in eq. \eqref{3}, one gets
\begin{eqnarray}\label{OUT_Partons}
\left(\frac{df(\mathbf{p})}{dt}\right)_{\rm out}=-f(\mathbf{p})\int\frac{d\mathbf{p^\prime}}{(2\pi)^3}\Gamma_{\mathbf{p},\mathbf{p^\prime}}\left(1-f(\mathbf{p^\prime})\right)
.\end{eqnarray}
In the similar way, the number of partons scattered into the volume element $d\mathbf{p}$ about $\mathbf{p}$ is expressed as
\begin{eqnarray}\label{IN_Partons}
\left(\frac{df(\mathbf{p})}{dt}\right)_{\rm in}=\left(1-f(\mathbf{p})\right)\int\frac{d\mathbf{p^\prime}}{(2\pi)^3}\Gamma_{\mathbf{p^\prime},\mathbf{p}}f(\mathbf{p^\prime})
.\end{eqnarray}
The evolution of the distribution function or the collision term is mainly contributed by the partons scattering in and out of the volume element due to collisions. Thus, we have
\begin{eqnarray}\label{Sum}
\nonumber\left(\frac{df(\mathbf{p})}{dt}\right)_{\rm coll} &=& \left(\frac{df(\mathbf{p})}{dt}\right)_{\rm out}+\left(\frac{df(\mathbf{p})}{dt}\right)_{\rm in} \\ &=& -\int\frac{d\mathbf{p^\prime}}{(2\pi)^3}\left[\Gamma_{\mathbf{p},\mathbf{p^\prime}}f(\mathbf{p})\left(1-f(\mathbf{p^\prime})\right)-\Gamma_{\mathbf{p^\prime},\mathbf{p}}f(\mathbf{p^\prime})\left(1-f(\mathbf{p})\right)\right]
,\end{eqnarray}
which is a nonlinear integro-differential Boltzmann transport equation. To find a solution of the Boltzmann transport equation, some simplifying approximations are needed. One of the frequently used simplifications for the collision term is known as the relaxation time approximation. This method is useful in linearizing the Boltzmann transport equation through the following assumptions: (i) The distribution function gets infinitesimally deviated from its equilibrium, so that within a phenomenological timescale (the relaxation time) $\tau$, the system returns back to the equilibrium state. (ii) The probability per unit time for a collision, {\em i.e.}, ${1}/{\tau(\mathbf{p})}$ no longer depends on the parton distribution function. (iii) The number of partons scattering into the phase space volume element, involve the equilibrium distribution function, whereas the number of partons which move out of the concerned phase space 
volume element after suffering collisions, involve the nonequilibrium 
distribution function. So, the collision term, {\em i.e.} the rate at which the distribution 
function changes due to collisions becomes
\begin{eqnarray}
\left(\frac{df(\mathbf{p})}{dt}\right)_{\rm coll}=-\frac{\left(f(p)-f_0(p)\right)}{\tau}
.\end{eqnarray}
Thus, the Boltzmann transport equation takes the linearized form via the relaxation time approximation as
\begin{eqnarray}
\frac{df(p)}{dt}=-\frac{\left(f(p)-f_0(p)\right)}{\tau}
.\end{eqnarray}
In the present work, we have considered the linearized Boltzmann transport equation in the relaxation time approximation method. 

It is known that the relaxation time approximation defies the local particle number conservation 
in the medium, because the charge is not conserved instantaneously but only on the average 
over a cycle. However, the validity of the conservation laws can be guaranteed by imposing a 
condition that the relaxation time has no momentum dependence and this can be perceived as follows. 
In general, the Boltzmann transport equation for a single particle distribution function $f(x,p)$ is 
written as
\begin{eqnarray}\label{Operator form}
p^\mu\partial_\mu f=\mathcal{C}[f]=\hat{L}\phi
,\end{eqnarray}
where $\hat{L}$ is the linearized collision operator, $f_0\phi=f-f_0=\delta f$ and $f_0$ denotes the equilibrium distribution function. The $\hat{L}$ has only the nonpositive eigenvalues, whose absolute values elucidate the reciprocal of the relaxation times of nonequilibrium perturbations and the eigenfunctions with zero eigenvalues are conserved in collisions, such as the particle number, energy and momentum \cite{Crecignani:2002,Rocha:PRL127'2021}. In the relaxation time approximation, eq. \eqref{Operator form} changes to take the following form, 
\begin{eqnarray}\label{R.T.A. form}
p^\mu\partial_\mu f=-\frac{\omega}{\tau}\delta f
,\end{eqnarray}
where $\omega=p_0$ is the energy of parton. In order to satisfy the fundamental conservation equations, the zeroth and the first moments of the collision integral must vanish, {\em i.e.} $\int dP ~ \mathcal{C}[f]=0$ and $\int dP ~ p^\mu \mathcal{C}[f]=0$, which follow from the particle number conservation and energy-momentum conservation, respectively. Thus, to check the conservation of the particle number, one needs to multiply both sides of eq. \eqref{R.T.A. form} with 1 and then integrate in momentum, {\em i.e.}, 
\begin{eqnarray}\label{P.N.C.}
\int dP p^\mu\partial_\mu f=-\int dP \frac{\omega}{\tau}\delta f
,\end{eqnarray}
where $\int dP=\int \frac{d^3{\rm p}}{(2\pi)^3p_0}$. Similarly, to check the conservation of the energy-momentum, one needs to multiply both sides of eq. \eqref{R.T.A. form} with $p^\nu$ and then integrate in momentum, {\em i.e.}, 
\begin{eqnarray}\label{E.M.C.}
\int dP p^\mu p^\nu\partial_\mu f=-\int dP p^\nu\frac{\omega}{\tau}\delta f
.\end{eqnarray}
Integrating and then simplifying, eq. \eqref{P.N.C.} and eq. \eqref{E.M.C.} turn out to be
\begin{eqnarray}\label{P.N.C. (1)}
&&\partial_\mu\left[\int dP p^\mu f\right]=-\int dP \frac{\omega}{\tau}\delta f, \\ 
&&\label{E.M.C. (1)}\partial_\mu\left[\int dP p^\mu p^\nu f\right]=-\int dP p^\nu\frac{\omega}{\tau}\delta f .\end{eqnarray}
If the relaxation time is independent of the momentum, then imposing the Landau matching conditions \cite{Landau:BOOK'1987}, one gets $\int dP \omega\delta f=0$ and $\int dP p^\nu\omega\delta f=0$. 
As a result, eq. \eqref{P.N.C. (1)} and eq. \eqref{E.M.C. (1)} become
\begin{eqnarray}\label{P.N.C. (2)}
&&\partial_\mu\left[\int dP p^\mu f\right]=0, \\ 
&&\label{E.M.C. (2)}\partial_\mu\left[\int dP p^\mu p^\nu f\right]=0
,\end{eqnarray}
which can be identified as the particle number conservation and the energy-momentum conservation, respectively. We have taken the relaxation time to be momentum-independent in this work. 

\subsection{Charge transport properties}
In the presence of an external electric field, the medium gets infinitesimally 
disturbed, and the spatial component of the induced electric current density 
can be expressed as
\begin{eqnarray}\label{current}
J^i=\sum_f g_f \int\frac{d^3\rm{p}}{(2\pi)^3}
\frac{p^i}{\omega_f} [q\delta f_f(x,p)+{\bar q}\delta \bar{f_f}(x,p)]
~,\end{eqnarray}
where `$f$' is the flavor index with $f=u,d,s$. In eq. \eqref{current}, 
$g_f$, $q$ ($\bar q$) and $\delta f_f $ 
($\delta \bar{f_f}$) are the degeneracy factor, the electric 
charge and the infinitesimal change in the distribution function for the 
quark (antiquark) of $f$th flavor, respectively. For a general 
configuration of electric and magnetic fields, the spatial component 
of the electric current density can be expressed as
\begin{eqnarray}\label{Multicomponent structure (1)}
J^i=\sigma^{ij}E_j=\sigma_0\delta^{ij}E_j+\sigma_1\epsilon^{ijk}b_kE_j+\sigma_2b^ib^jE_j
~,\end{eqnarray}
where $\sigma_0$, $\sigma_1$ and $\sigma_2$ are various charge transport coefficients in the presence of magnetic field and $\mathbf{b}=\frac{\mathbf{B}}{B}$ represents the direction of 
magnetic field. In eq. \eqref{Multicomponent structure (1)}, if we consider the case where the electric field and the magnetic field are perpendicular to each other, then the third term will vanish. Thus, eq. \eqref{Multicomponent structure (1)} can be rewritten as
\begin{eqnarray}\label{Multicomponent structure}
J^i=\sigma^{ij}E_j=\left(\sigma_{\rm el}\delta^{ij}+\sigma_{\rm H}\epsilon^{ij}\right)E_j
~,\end{eqnarray}
where $\epsilon^{ij}$ denotes the antisymmetric $2\times2$ unit matrix and one can identify $\sigma_0$ as electrical conductivity ($\sigma_{\rm el}$) and $\sigma_1$ as Hall conductivity ($\sigma_{\rm H}$). 

It is possible to obtain the electrical and Hall conductivities by 
comparing eq. \eqref{current} and 
eq. \eqref{Multicomponent structure}. The nonequilibrium part of the 
distribution function, {\em i.e.} $\delta f_f$ can be calculated from the 
relativistic Boltzmann transport equation, which has the following form 
in the relaxation time approximation \cite{Crecignani:2002}, 
\be\label{R.B.T.E.}
p^\mu\frac{\partial f_f(x,p)}{\partial x^\mu}+\mathcal{F}^\mu\frac{\partial f_f(x,p)}{\partial p^\mu}=-\frac{p_\nu u^\nu}{\tau_f}\delta f_f(x,p)
~,\ee
where $u^\nu$ denotes the four-velocity of fluid, $f_f=\delta f_f+f_f^0$, $\mathcal{F}^\mu=qF^{\mu\nu}p_\nu=(p^0\mathbf{v}\cdot\mathbf{F}, p^0\mathbf{F})$ with $F^{\mu\nu}$ being the electromagnetic field strength tensor. The Lorentz force 
is defined as $\mathbf{F}=q(\mathbf{E}+\mathbf{v}\times\mathbf{B})$. The 
components of $F^{\mu\nu}$ are related to the components of electric and 
magnetic fields as $F^{0i}=E^i$, $F^{i0}=-E^i$ 
and $F^{ij}=\frac{1}{2}\epsilon^{ijk}B_k$. In eq. \eqref{R.B.T.E.}, $\tau_f$ 
represents the relaxation time of a quark with flavor $f$. According to the 
assumption of the relaxation time approximation, the system gets slightly 
deviated from equilibrium due to the action of the external perturbation 
and $\tau_f$ defines the time required by a nonequilibrium system to 
return back to its equilibrium state. The relaxation time considered 
is the mean relaxation time and thus $\tau_f$ does not depend on energy and 
momentum. In addition, for weak magnetic field regime, the magnetic field is not the dominant scale as compared to the temperature scale of the thermal system in equilibrium. So, the effects of Landau quantization on the phase space and on the scattering processes have not been considered in the present work. In this regime, the dependence of magnetic field and chemical potential in $\tau_f$ enters through the running coupling constant only (unlike in the strong magnetic field regime, where the presence of strong magnetic field restricts the motion of charged particles to only one spatial dimension, thus severely affecting the relaxation time \cite{Hattori:PRD95'2017,Hattori:PRD96'2017,Kurian:PRD97'2018}). The relaxation time for quarks (antiquarks), $\tau_f$ ($\tau_{\bar{f}}$) in a thermal medium is given \cite{Hosoya:NPB250'1985} by
\begin{eqnarray}
\tau_{f(\bar{f})}=\frac{1}{5.1T\alpha_s^2\log\left(1/\alpha_s\right)\left[1+0.12(2N_f+1)\right]}
~.\end{eqnarray}
Here, the dependence of magnetic field and chemical potential enters through the 
running coupling constant ($\alpha_s$) \cite{Ayala:PRD98'2018}, 
\be\label{R.C.}
\alpha_s\left(\Lambda^2, eB\right)=\frac{\alpha_s\left(\Lambda^2\right)}{1+b_1\alpha_s\left(\Lambda^2\right)\ln\left(\frac{\Lambda^2}{\Lambda^2+eB}\right)}
,\ee
where $\alpha_s\left(\Lambda^2\right)$ is given by
\be\label{R.C.1}
\alpha_s\left(\Lambda^2\right)=\frac{1}{b_1\ln\left(\frac{\Lambda^2}{\Lambda_{\rm\overline{MS}}^2}\right)}
,\ee
with $b_1=\frac{11N_c-2N_f}{12\pi}$, $\Lambda_{\rm\overline{MS}}=0.176$ GeV 
and $\Lambda=2\pi\sqrt{T^2+\mu_f^2/\pi^2}$ for quarks and antiquarks. The 
equilibrium distribution functions for quark and antiquark of $f$th flavor 
are written as
\be\label{D.F.}
&&f_f^0=\frac{1}{e^{\beta(\omega_f-\mu_f)}+1}~, \\ 
&&\label{D.F.1}\bar{f_f}^0=\frac{1}{e^{\beta(\omega_f+\mu_f)}+1}
~,\ee
respectively, where $\omega_f=\sqrt{\mathbf{p}^2+m_f^2}$, $T=\beta^{-1}$ and 
$\mu_f$ is the chemical potential of $f$th flavor of quark. The relativistic 
Boltzmann transport equation \eqref{R.B.T.E.} can be rewritten as
\be\label{R.B.T.E.(1)}
\frac{\partial f_f}{\partial t}+\mathbf{v}\cdot\frac{\partial f_f}{\partial \mathbf{r}}+\frac{\mathbf{p}\cdot\mathbf{F}}{p_0}\frac{\partial f_f}{\partial p_0}+\mathbf{F}\cdot\frac{\partial f_f}{\partial \mathbf{p}}=-\frac{(f_f-f_f^0)}{\tau_f}
~.\ee
In the case of a spatially homogeneous distribution function and 
for the steady-state condition, we can take 
$\frac{\partial f_f}{\partial \mathbf{r}}=0$ and 
$\frac{\partial f_f}{\partial t}=0$. Thus, eq. \eqref{R.B.T.E.(1)} turns out to be 
\be\label{R.B.T.E.(2)}
\mathbf{v}\cdot\mathbf{F}\frac{\partial f_f}{\partial p_0}+\mathbf{F}\cdot\frac{\partial f_f}{\partial \mathbf{p}}=-\frac{(f_f-f_f^0)}{\tau_f}
~.\ee
For an electric field along x-direction ($\mathbf{E}=E\hat{x}$) and a magnetic 
field along z-direction ($\mathbf{B}=B\hat{z}$), we get
\be\label{R.B.T.E.(3)}
\tau_fqEv_x\frac{\partial f_f}{\partial p_0}+\tau_fqBv_y\frac{\partial f_f}{\partial p_x}-\tau_fqBv_x\frac{\partial f_f}{\partial p_y}=f_f^0-f_f-\tau_fqE\frac{\partial f_f^0}{\partial p_x}
~.\ee
In order to solve the above equation, we have used the following ansatz which was first suggested by ref. \cite{Feng:PRD96'2017}, 
\be\label{ansatz}
f_f=f_f^0-\tau_fq\mathbf{E}\cdot\frac{\partial f_f^0}{\partial \mathbf{p}}-\mathbf{\Gamma}\cdot\frac{\partial f_f^0}{\partial \mathbf{p}}
~.\ee
This ansatz is formulated in such a way that it depends on both electric and magnetic fields and is relevant in the weak magnetic field limit. In this limit, quantities can be expanded in powers of $eB$ and terms with higher orders of $eB$ can be neglected. The above ansatz also satisfies this weak magnetic field condition, because in eq. \eqref{ansatz}, first term is the equilibrium distribution function, second term is of the order $\mathcal{O}((eB)^0)$ and third term is of the order $\mathcal{O}((eB)^1)$. Thus, the unknown quantity $\mathbf{\Gamma}$ in eq. \eqref{ansatz} requires to be related to the magnetic field, {\em i.e.} it should depend on $eB$. 

Assuming the quark distribution function to be much closer to equilibrium, we have
\begin{eqnarray*}
&& \frac{\partial f_f^0}{\partial p_x}=-\beta v_xf_f^0\left(1-f_f^0\right), ~~ \frac{\partial f_f^0}{\partial p_y}=-\beta v_yf_f^0\left(1-f_f^0\right), ~~ \frac{\partial f_f^0}{\partial p_z}=-\beta v_zf_f^0\left(1-f_f^0\right)
~.\end{eqnarray*}
Thus using the above relations and the ansatz \eqref{ansatz}, eq. \eqref{R.B.T.E.(3)} can be simplified at high temperature as
\be\label{R.B.T.E.(4)}
\tau_fqEv_x\frac{\partial f_f}{\partial p_0}+\beta f_f^0\left(\Gamma_xv_x+\Gamma_yv_y+\Gamma_zv_z\right)-qB\tau_f\left(v_x\frac{\partial f_f}{\partial p_y}-v_y\frac{\partial f_f}{\partial p_x}\right)=0
~.\ee
From eq. \eqref{R.B.T.E.(4)} and ansatz \eqref{ansatz}, we get the 
infinitesimal change of the quark distribution function 
(calculated in appendix \ref{I.C. of Q.D.F.1}) as
\be\label{deltaf.q}
\delta f_f=2qEv_x\beta\left(\frac{\tau_f}{1+\omega_c^2\tau_f^2}\right)f_f^0\left(1-f_f^0\right)-2qEv_y\beta\left(\frac{\omega_c\tau_f^2}{1+\omega_c^2\tau_f^2}\right)f_f^0\left(1-f_f^0\right)
.\ee
Similarly for antiquarks, we get
\be\label{deltaf.aq}
\delta \bar{f_f}=2\bar{q}Ev_x\beta\left(\frac{\tau_{\bar{f}}}{1+\omega_c^2\tau_{\bar{f}}^2}\right)\bar{f_f}^0\left(1-\bar{f_f}^0\right)
-2\bar{q}Ev_y\beta\left(\frac{\omega_c\tau_{\bar{f}}^2}{1+\omega_c^2\tau_{\bar{f}}^2}\right)\bar{f_f}^0\left(1-\bar{f_f}^0\right)
.\ee
Substituting the values of $\delta f_f$ and $\delta \bar{f_f}$ in eq. \eqref{current} and then comparing with eq. \eqref{Multicomponent structure}, we get the electrical conductivity and the Hall conductivity for a dense QCD medium in a weak magnetic field as
\be\label{E.C.}
&&\sigma_{\rm el}=\frac{\beta}{3\pi^2}\sum_f g_f q_f^2\int d{\rm p}~\frac{{\rm p}^4}{\omega_f^2} ~ \left[\frac{\tau_f}{1+\omega_c^2\tau_f^2}f_f^0\left(1-f_f^0\right)+\frac{\tau_{\bar{f}}}{1+\omega_c^2\tau_{\bar{f}}^2}\bar{f_f}^0\left(1-\bar{f_f}^0\right)\right], \\ 
&&\label{H.Conductivity}\sigma_{\rm H}=\frac{\beta}{3\pi^2}\sum_f g_f q_f^2\int d{\rm p}~\frac{{\rm p}^4}{\omega_f^2} ~ \left[\frac{\omega_c\tau_f^2}{1+\omega_c^2\tau_f^2}f_f^0\left(1-f_f^0\right)+\frac{\omega_c\tau_{\bar{f}}^2}{1+\omega_c^2\tau_{\bar{f}}^2}\bar{f_f}^0\left(1-\bar{f_f}^0\right)\right]
.\ee

\subsection{Heat transport properties}
Due to the presence of a temperature gradient, the system gets 
deviated from its equilibrium state, resulting in heat 
flow. This heat flow is directly proportional to the temperature 
gradient with the proportionality factor being the thermal 
conductivity. The study of thermal conductivity can shed light on the 
understanding of the heat transport in the medium and its possible 
effect on the hydrodynamic equilibrium of the medium. 

The heat flow four-vector is defined by the difference between the 
energy diffusion and the enthalpy diffusion, 
\be\label{heat flow}
Q_\mu=\Delta_{\mu\alpha}T^{\alpha\beta}u_\beta-h\Delta_{\mu\alpha}N^\alpha
.\ee
Here, the projection operator $\Delta_{\mu\alpha}=g_{\mu\alpha}-u_\mu u_\alpha$ 
and the enthalpy per particle $h=(\varepsilon+P)/n$ 
with the particle number density $n=N^\alpha u_\alpha$, the 
energy density $\varepsilon=u_\alpha T^{\alpha\beta} u_\beta$ and the 
pressure $P=-\Delta_{\alpha\beta}T^{\alpha\beta}/3$. The particle flow four-vector $N^\alpha$ and the energy-momentum tensor $T^{\alpha\beta}$ are respectively defined as
\be
&&N^\alpha=\sum_f g_f\int \frac{d^3{\rm p}}{(2\pi)^3\omega_f}p^\alpha \left[f_f(x,p)+\bar{f}_f(x,p)\right] \label{P.F.F.}, \\ &&T^{\alpha\beta}=\sum_f g_f\int \frac{d^3{\rm p}}{(2\pi)^3\omega_f}p^\alpha p^\beta \left[f_f(x,p)+\bar{f}_f(x,p)\right] \label{E.M.T.}
.\ee
In the rest frame of the heat bath, $Q_\mu u^\mu=0$, so the heat flow is spatial, 
which is given by
\be\label{heat1}
Q^i=\sum_f g_f\int \frac{d^3{\rm p}}{(2\pi)^3} ~ \frac{p^i}{\omega_f}\left[(\omega_f-h_f)\delta f_f(x,p)+(\omega_f-\bar{h}_f)\delta \bar{f}_f(x,p)\right]
.\ee
Through the Navier-Stokes equation, the heat flow is related to the gradients 
of temperature and pressure \cite{Greif:PRE87'2013} as
\be\label{heat}
Q^i &=& -\kappa^{ij}\left[\partial_j T - \frac{T}{\varepsilon+P}\partial_j P\right] 
.\ee
At finite magnetic field, $Q^i$ is expressed as
\begin{eqnarray}\label{Multicomponent structure (2)}
Q^i=-\left(\kappa_0\delta^{ij}+\kappa_1\epsilon^{ijk}b_k+\kappa_2b^ib^j\right)
\left[\partial_j T - \frac{T}{\varepsilon+P}\partial_j P\right] 
,\end{eqnarray}
where $\kappa_0$, $\kappa_1$ and $\kappa_2$ are various heat transport coefficients 
in the presence of magnetic field and $\mathbf{b}=\frac{\mathbf{B}}{B}$. In eq. \eqref{Multicomponent structure (2)}, if we consider the case where the gradients 
of temperature and pressure are perpendicular to the magnetic field, then the third term 
will vanish. Thus, eq. \eqref{Multicomponent structure (2)} turns out to be 
\begin{eqnarray}\label{heat2}
Q^i=-\left(\kappa_0\delta^{ij}+\kappa_1\epsilon^{ij}\right)\left[\partial_j T - \frac{T}{\varepsilon+P}\partial_j P\right] 
.\end{eqnarray}
By comparing equations (\ref{heat1}) and (\ref{heat2}), one can 
obtain thermal conductivity ($\kappa_0$) and Hall-type thermal 
conductivity ($\kappa_1$). With the help of the 
ansatz \eqref{ansatz}, the relativistic Boltzmann transport 
equation \eqref{R.B.T.E.} can be rewritten as
\be\label{eq.1}
\frac{\tau_f}{p_0}p^\mu\frac{\partial f_f^0}{\partial x^\mu}+\beta f_f^0\left(\Gamma_xv_x+\Gamma_yv_y+\Gamma_zv_z\right)+\tau_fqEv_x\frac{\partial f_f}{\partial p_0}-qB\tau_f\left(v_x\frac{\partial f_f}{\partial p_y}-v_y\frac{\partial f_f}{\partial p_x}\right)=0
.\ee
Since magnetic field is taken along z-direction, no explicit dependence of magnetic field
on spatial derivative of the distribution function along z-direction can be observed. By 
comparing both sides of eq. \eqref{eq.1}, one gets $\Gamma_z=0$. Substituting the values 
of $\frac{\partial f_f}{\partial p_0}$, $\frac{\partial f_f}{\partial p_x}$ and 
$\frac{\partial f_f}{\partial p_y}$ in the above equation and then 
dropping higher order velocity terms, we have
\be\label{eq.2}
L-\beta f_f^0\tau_fqEv_x+\beta f_f^0\left(\Gamma_xv_x+\Gamma_yv_y\right)-\frac{qB\tau_f\beta f_f^0}{\omega_f}\left(v_x\Gamma_y-v_y\Gamma_x\right)+\frac{\tau_f^2qBqEv_y\beta f_f^0}{\omega_f}=0
,\ee
where $L=\frac{\tau_f}{p_0}p^\mu\frac{\partial f_f^0}{\partial x^\mu}$. For quark distribution function, $L$ is calculated as
\be\label{L}
\nonumber L &=& \tau_f\beta f_f^0\frac{\left(\omega_f-h_f\right)}{T}v_x\left(\partial^xT-\frac{T}{nh_f}\partial^xP\right)+\tau_f\beta f_f^0\frac{\left(\omega_f-h_f\right)}{T}v_y\left(\partial^yT-\frac{T}{nh_f}\partial^yP\right) \\ && +\tau_f\beta f_f^0\left[p_0\frac{DT}{T}-\frac{p^\mu p^\alpha}{p_0}\nabla_\mu u_\alpha+TD\left(\frac{\mu_f}{T}\right)\right]
.\ee
With the help of eq. \eqref{eq.2}, eq. \eqref{L} and ansatz \eqref{ansatz}, 
we get the infinitesimal change of the quark distribution function 
(calculated in appendix \ref{I.C. of Q.D.F.2}) as
\be\label{deltaf.q1}
\nonumber\delta f_f &=& \frac{2qE\tau_fv_x\beta f_f^0\left(1-f_f^0\right)}{1+\omega_c^2\tau_f^2}-\frac{2qE\omega_c\tau_f^2v_y\beta f_f^0\left(1-f_f^0\right)}{1+\omega_c^2\tau_f^2}-\beta^2 f_f^0\left(1-f_f^0\right)\frac{\tau_f(\omega_f-h_f)}{\left(1+\omega_c^2\tau_f^2\right)} \\ && \nonumber\times\left[v_x\left(\partial^xT-\frac{T}{nh_f}\partial^xP\right)+v_y\left(\partial^yT-\frac{T}{nh_f}\partial^yP\right)\right]-\beta^2 f_f^0\left(1-f_f^0\right) \\ && \times\frac{\omega_c\tau_f^2(\omega_f-h_f)}{\left(1+\omega_c^2\tau_f^2\right)}\left[v_x\left(\partial^yT-\frac{T}{nh_f}\partial^yP\right)-v_y\left(\partial^xT-\frac{T}{nh_f}\partial^xP\right)\right]
.\ee
Similarly, the infinitesimal change of the antiquark 
distribution function is calculated as
\be\label{deltaf.aq1}
\nonumber\delta \bar{f_f} &=& \frac{2\bar{q}E\tau_{\bar{f}}v_x\beta \bar{f_f}^0\left(1-\bar{f_f}^0\right)}{1+\omega_c^2\tau_{\bar{f}}^2}-\frac{2\bar{q}E\omega_c\tau_{\bar{f}}^2v_y\beta \bar{f_f}^0\left(1-\bar{f_f}^0\right)}{1+\omega_c^2\tau_{\bar{f}}^2}-\beta^2 \bar{f_f}^0\left(1-\bar{f_f}^0\right)\frac{\tau_{\bar{f}}(\omega_f-\bar{h}_f)}{\left(1+\omega_c^2\tau_{\bar{f}}^2\right)} \\ && \nonumber\times\left[v_x\left(\partial^xT-\frac{T}{n\bar{h}_f}\partial^xP\right)+v_y\left(\partial^yT-\frac{T}{n\bar{h}_f}\partial^yP\right)\right]-\beta^2 \bar{f_f}^0\left(1-\bar{f_f}^0\right) \\ && \times\frac{\omega_c\tau_{\bar{f}}^2(\omega_f-\bar{h}_f)}{\left(1+\omega_c^2\tau_{\bar{f}}^2\right)}\left[v_x\left(\partial^yT-\frac{T}{n\bar{h}_f}\partial^yP\right)-v_y\left(\partial^xT-\frac{T}{n\bar{h}_f}\partial^xP\right)\right]
.\ee
Substituting $\delta f_f$ and $\delta \bar{f_f}$ in eq. \eqref{heat1} 
and then comparing with eq. \eqref{heat2}, we get the thermal 
conductivity and the Hall-type thermal conductivity for a dense QCD 
medium in a weak magnetic field as
\be\label{H.C.}
\nonumber\kappa_0 &=& \frac{\beta^2}{6\pi^2}\sum_f g_f\int d{\rm p}~\frac{{\rm p}^4}{\omega_f^2} ~ \left[\frac{\tau_f}{1+\omega_c^2\tau_f^2}\left(\omega_f-h_f\right)^2f_f^0\left(1-f_f^0\right)\right. \\ && \left.+\frac{\tau_{\bar{f}}}{1+\omega_c^2\tau_{\bar{f}}^2}\left(\omega_f-\bar{h}_f\right)^2
\bar{f_f}^0\left(1-\bar{f_f}^0\right)\right], \\ 
\label{H.C.(1)}\nonumber\kappa_1 &=& \frac{\beta^2}{6\pi^2}\sum_f g_f\int d{\rm p}~\frac{{\rm p}^4}{\omega_f^2} ~ \left[\frac{\omega_c\tau_f^2}{1+\omega_c^2\tau_f^2}\left(\omega_f-h_f\right)^2f_f^0\left(1-f_f^0\right)\right. \\ && \left.+\frac{\omega_c\tau_{\bar{f}}^2}{1+\omega_c^2\tau_{\bar{f}}^2}\left(\omega_f-\bar{h}_f\right)^2
\bar{f_f}^0\left(1-\bar{f_f}^0\right)\right]
.\ee

\section{Lifetime of magnetic field}
In relativistic heavy ion collisions, extremely strong magnetic fields are 
produced in a direction perpendicular to the collision plane. These 
magnetic fields are transient in nature and become weak with time. However, the 
presence of electrical conductivity in the medium plays a crucial role in 
significantly extending the lifetimes of such fields. In addition, the finite chemical potential of  the medium might affect the lifetime of magnetic field. Thus, the study of the 
variation of magnetic field with time for an electrically conducting medium at finite chemical potential ($\mu$) is relevant to this work. 

\begin{figure}[]
\begin{center}
\includegraphics[width=7.4cm]{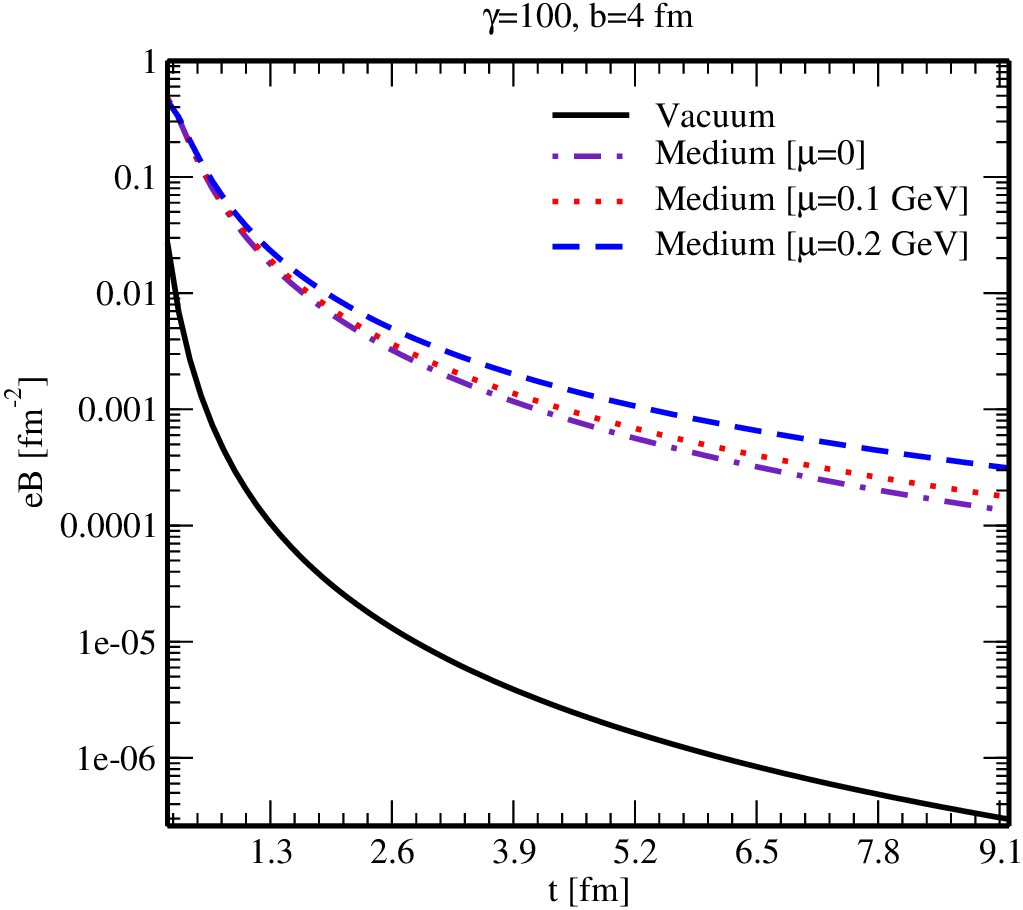}
\caption{The variation of magnetic field with time in vacuum and in a thermally 
conducting medium at different chemical potentials for impact parameter 
$b=4$ fm and Lorentz factor $\gamma=100$.}\label{eb}
\end{center}
\end{figure}

Consider a charged particle moving along $x$-direction. According to 
Maxwell's equations, a magnetic field will be created in a direction 
perpendicular to the particle trajectory, and can be expressed 
\cite{Tuchin:AHEP2013'2013} as
\be\label{eb1}
e\mathbf{B}_{\rm medium}=\frac{e^2b\sigma_{\rm el}}{8\pi(t-x)^2}e^{-\frac{b^2\sigma_{\rm el}}{4(t-x)}}\hat{\mathbf{z}}
~,\ee
whereas, the magnetic field produced in vacuum is 
expressed \cite{Tuchin:AHEP2013'2013} as
\be\label{ebv}
e\mathbf{B}_{\rm vacuum}=\frac{e^2b\gamma}{4\pi\left\lbrace b^2+\gamma^2(t-x)^2 \right\rbrace^{3/2}}\hat{\mathbf{z}}
~.\ee
Here, $b$ and $\gamma$ are the impact parameter and the 
Lorentz factor of heavy ion collision, respectively. In 
eq. \eqref{eb1}, the electrical conductivity has been taken 
as a function of the time through the cooling law, $T^3\propto{t^{-1}}$. 
The initial time and temperature are fixed at $0.2$ fm and $390$ MeV, 
respectively. Figure \ref{eb} shows the variation of magnetic field with time in 
vacuum and in a thermal medium at different chemical potentials for $x=0$, 
$b=4$ fm, $\gamma=100$ and $\sqrt{s}=200$ GeV. 

\begin{figure}[]
\begin{center}
\includegraphics[width=7.4cm]{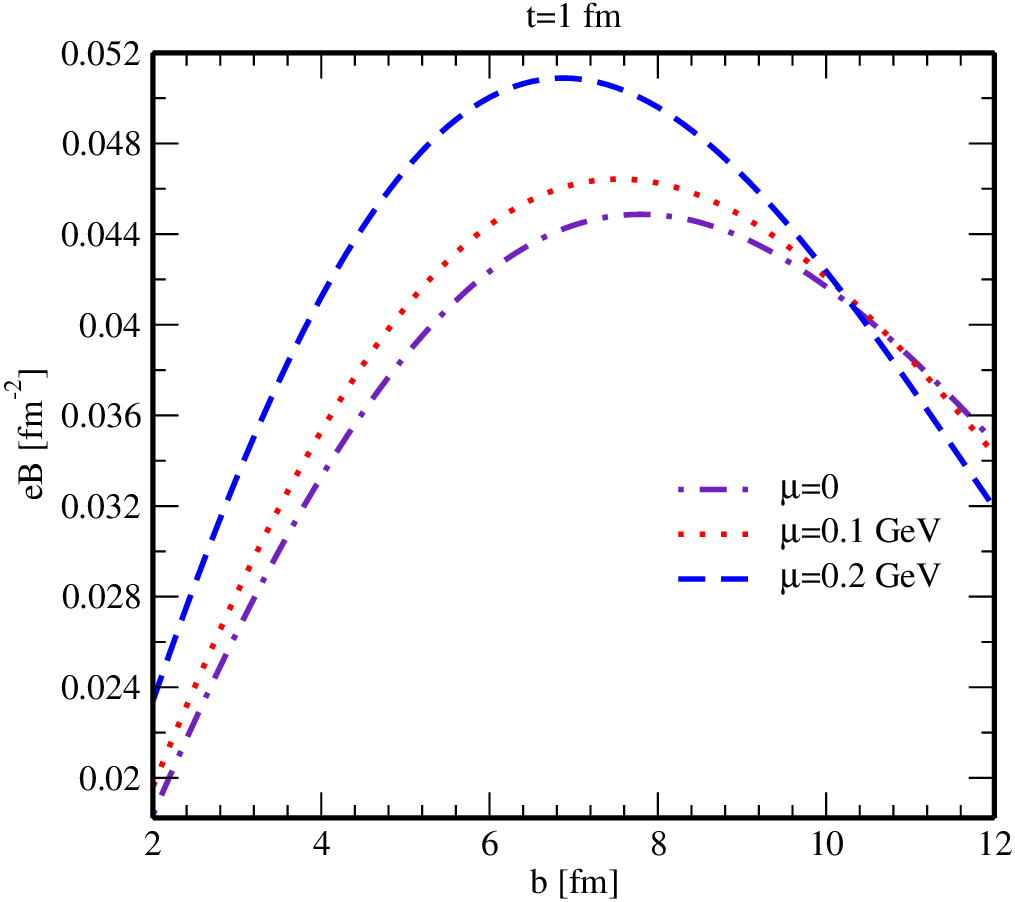}
\caption{The variation of magnetic field with impact parameter in a thermally 
conducting medium for different values of chemical 
potential at time $t=1$ fm.}\label{eb.i.p.}
\end{center}
\end{figure}
It can be observed that the strength of magnetic field decays 
very fast in the vacuum, whereas in an electrically conducting 
medium, its decay becomes much slower. Initially, the decrease 
in the strength of magnetic field in the thermal 
medium is noticeably high, however it gradually saturates with the time, 
which explains that, as compared to the strong 
magnetic field, the weak magnetic field can stay longer. In figure \ref{eb}, 
we have also noticed that the presence of chemical potential in the medium 
helps in elongating the lifetime of magnetic field. Thus, the properties 
of a dense thermal medium are expected to be influenced by the magnetic 
field. However at initial time, the difference between the variations of 
magnetic field in two mediums at zero chemical potential and finite chemical 
potential is less conspicuous. 

Figure \ref{eb.i.p.} displays the variation of the strength of magnetic field 
with impact parameter at a fixed time $t=1$ fm for different values of 
chemical potential. It can be observed that the trend of variation of $eB$ 
with $b$ is not monotonic. It increases with $b$ peaking at $b$ values 
for mid-central collision and then it shows a gradual decrease. With the increase 
of chemical potential, the peak shifts towards lower $b$ values. Thus, for a 
larger chemical potential, the magnetic field attains its highest strength at a 
smaller impact parameter. 

\section{Results and discussions}
In this section, we are going to discuss the results on different charge and heat transport coefficients by using the thermal masses of charged particles in the quasiparticle description. It should be noted that the relaxation time approximation of the Boltzmann transport equation does not include the interactions among the constituents of the medium. However, in the quasiparticle description, each parton acquires a quasipaticle/thermally generated mass, which basically incorporates the interactions of the concerned parton with other partons in the medium. In this description, the QGP medium is described as a medium consisting of thermally massive noninteracting quasipartons. The quasiparticle masses predominantly depend on temperature 
of the medium and for a dense thermal medium, they depend on chemical 
potential ($\mu_f$) too. The quasiparticle mass (squared) of the $f$th parton is given \cite{Braaten:PRD45'1992,Peshier:PRD66'2002} by
\be\label{Q.P.M.}
m_{fT}^2=\frac{g^2T^2}{6}\left(1+\frac{\mu_f^2}{\pi^2T^2}\right)
,\ee
with $g^2=4\pi\alpha_s$, where $\alpha_s$ represents the one-loop strong 
running coupling constant at finite temperature, chemical potential and magnetic 
field and is defined in eq. \eqref{R.C.}. Here the magnetic 
field-dependence enters only through $\alpha_s$, at least in the weak 
magnetic field regime. The chemical potentials for all quark flavors are kept 
same, {\em i.e.} $\mu_f=\mu$. 

The quasiparticle description does not change the form of the Boltzmann transport equation, however, it does change the equilibrium distribution function via the dispersion relation. In the Boltzmann transport equation \eqref{R.B.T.E.(2)}, the term involving the electromagnetic/Lorentz force is $\mathbf{F}\cdot\left[\mathbf{v}\frac{\partial f_f}{\partial p_0}+\frac{\partial f_f}{\partial \mathbf{p}}\right]$, where the factors $\frac{\partial f_f}{\partial p_0}$ and $\frac{\partial f_f}{\partial \mathbf{p}}$ get affected by the quasiparticle description. It is important to note that the quasiparticle model describes the mutual interactions of the partons in the thermal medium, but not their response to an external force field, so, the Lorentz force ($\mathbf{F}$) remains unaltered. Thus, the overall effect of quasiparticle mass on the Boltzmann transport equation is encoded in the factors $\frac{\partial f_f}{\partial p_0}$ and $\frac{\partial f_f}{\partial \mathbf{p}}$. On the other hand, in effective mass models \cite{Dusling:PRC85'2012,Alqahtani:PRC95'2017}, the terms concerning the mean field contribution had been added to the kinetic theory definition of the energy-momentum tensor. Similarly, the mean field term of the effective fugacity model of kinetic theory involves the fugacity parameters and their derivatives, and ref. \cite{Kurian:PRD99'2019} had reported that the mean field effects are negligible at high temperatures due to very slow variation of the effective fugacity with temperature. This suggests that, unlike effective mass models in which one needs to add terms concerning the mean field contribution to the kinetic theory \cite{Dusling:PRC85'2012,Alqahtani:PRC95'2017,Kurian:PRD99'2019}, in our case of quasiparticle 
description, no additional mean field term is required, as the modified dispersions of the 
quasiparticles take care of the thermal mass effect in the Boltzmann transport equation of kinetic theory. 

\begin{figure}[]
\begin{center}
\begin{tabular}{c c}
\includegraphics[width=6.86cm]{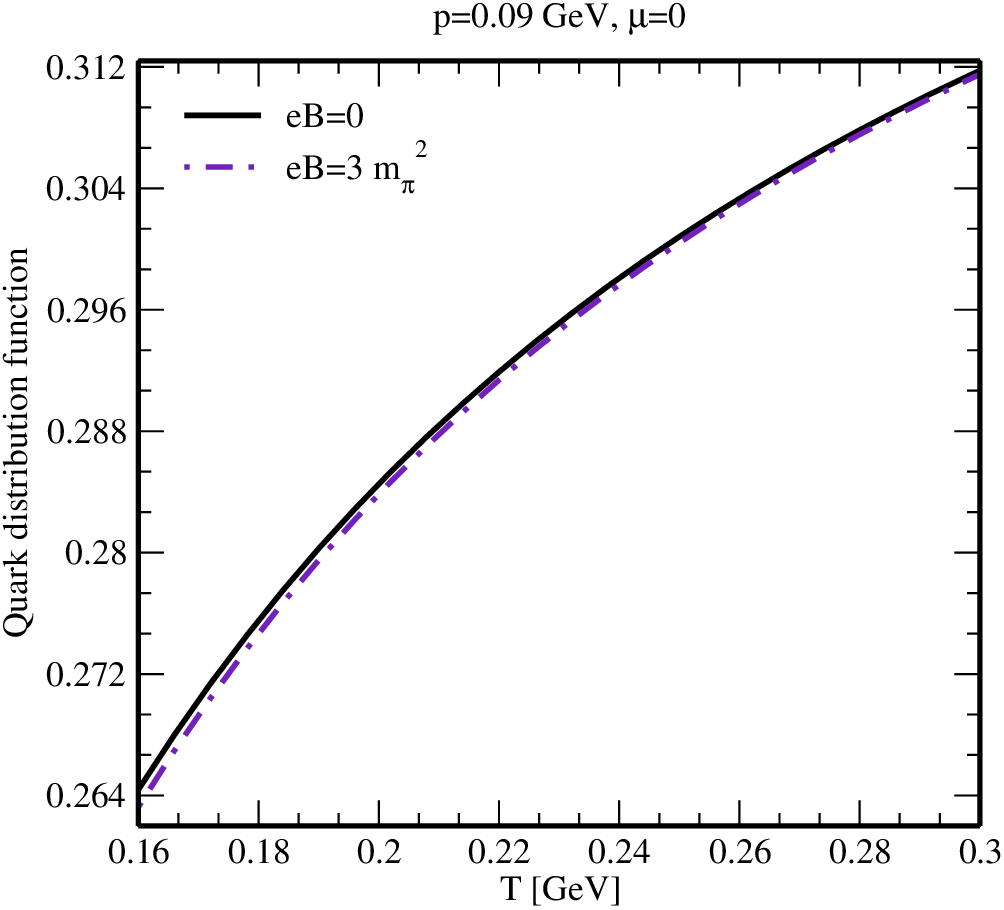}&
\hspace{0.423 cm}
\includegraphics[width=6.86cm]{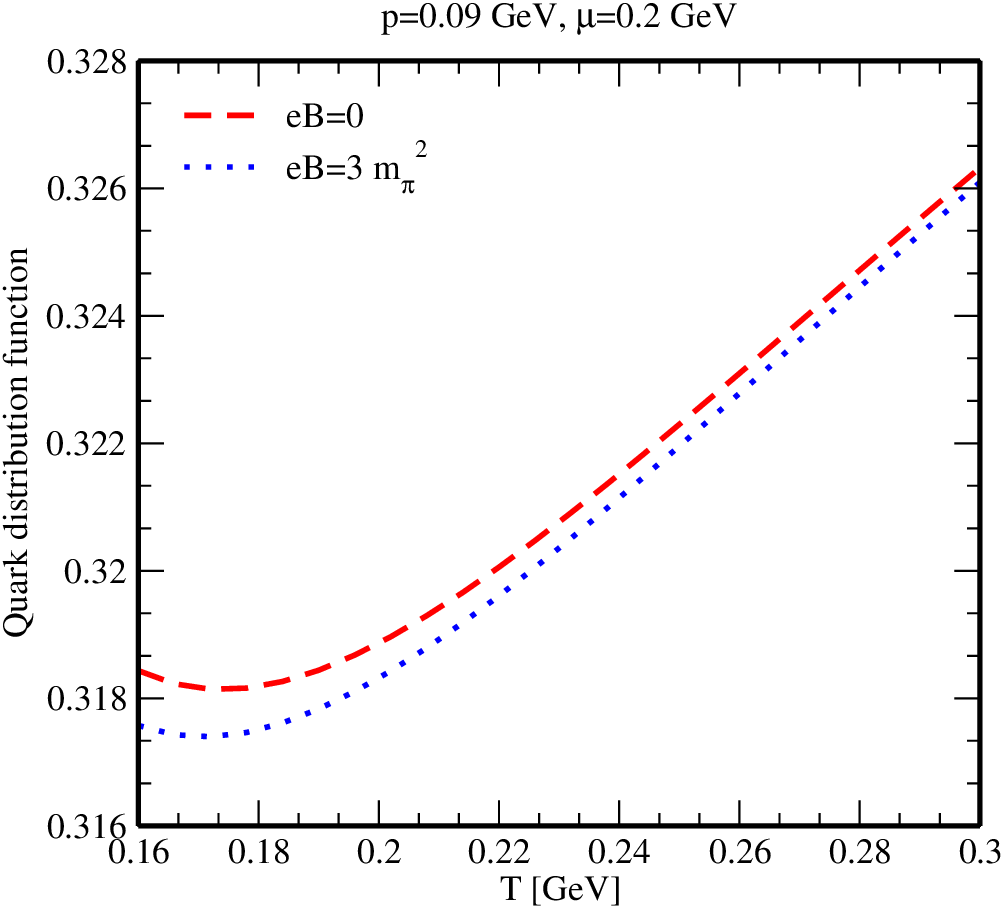} \\
a & b
\end{tabular}
\caption{The quark distribution function as a function of temperature for a fixed momentum.}\label{dftu}
\end{center}
\end{figure}

\begin{figure}[]
\begin{center}
\begin{tabular}{c c}
\includegraphics[width=6.86cm]{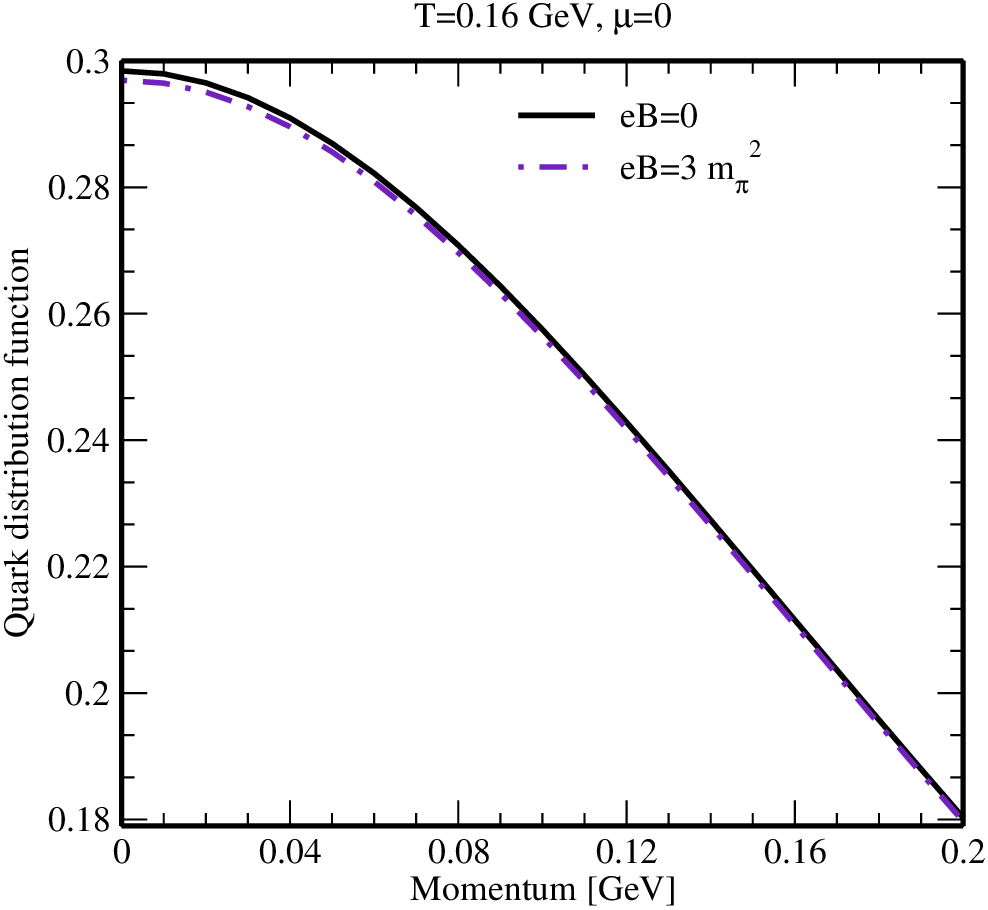}&
\hspace{0.423 cm}
\includegraphics[width=6.86cm]{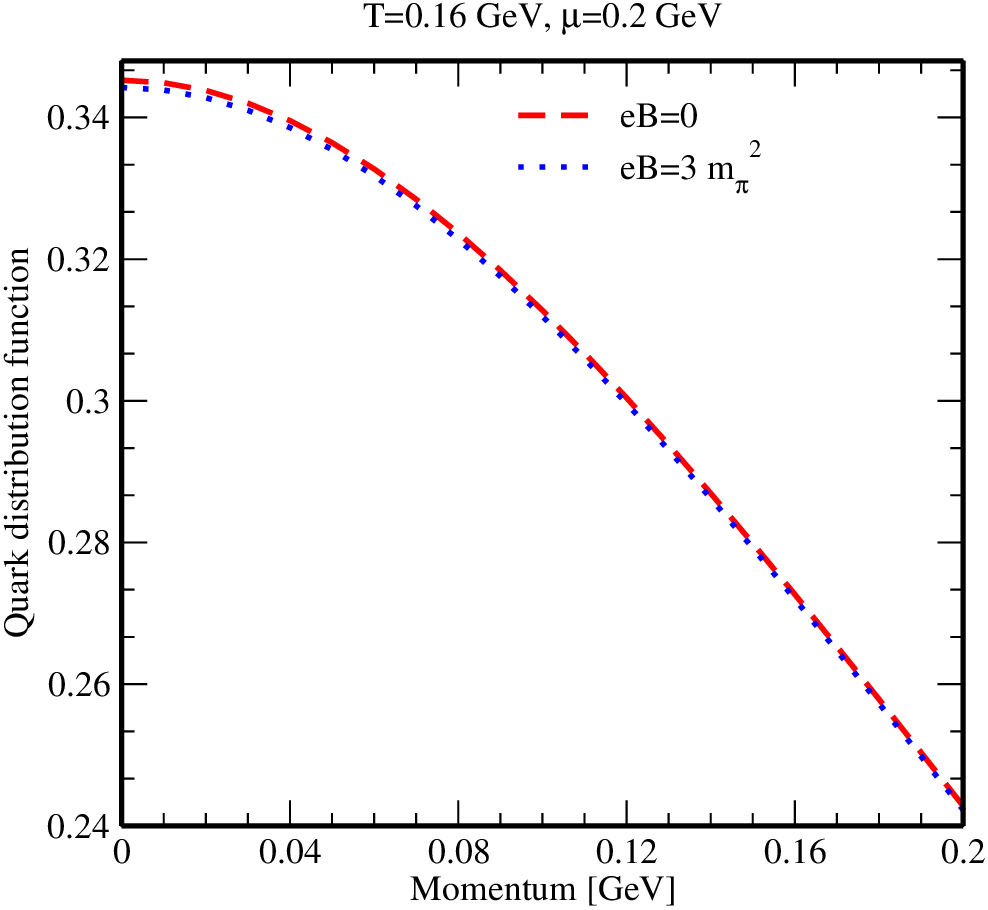} \\
a & b
\end{tabular}
\caption{The quark distribution function 
as a function of  momentum at a fixed temperature.}\label{dfpu}
\end{center}
\end{figure}

In the kinetic theory, the transport coefficients and their relative 
behavior mostly depend on the distribution function. So, before 
discussing the results on different transport coefficients, let us 
observe how the distribution function of $u$ quark varies with temperature 
and with momentum in the presence of weak magnetic field and finite 
chemical potential. The quark distribution function has been shown as a 
function of temperature in figure \ref{dftu} and as a function of 
momentum in figure \ref{dfpu}. It can be seen that the 
distribution function gets slightly decreased in the presence 
of a weak magnetic field in comparison to that in the pure thermal 
medium at zero magnetic field. However, a finite value of chemical 
potential significantly affects the magnitude and the shape of the 
distribution function (figures \ref{dftu}b and \ref{dfpu}b) as 
compared to the zero chemical potential case (figures \ref{dftu}a and \ref{dfpu}a). 
In the low temperature regime, the difference between the distribution functions at 
zero chemical potential and at finite chemical potential is larger. However, at 
high temperature, they tend to approach each other, as $\frac{\mu}{T}$ gets waned 
with the increase of temperature. 

\subsection{Components of charge transport}
\begin{figure}[]
\begin{center}
\begin{tabular}{c c}
\includegraphics[width=6.86cm]{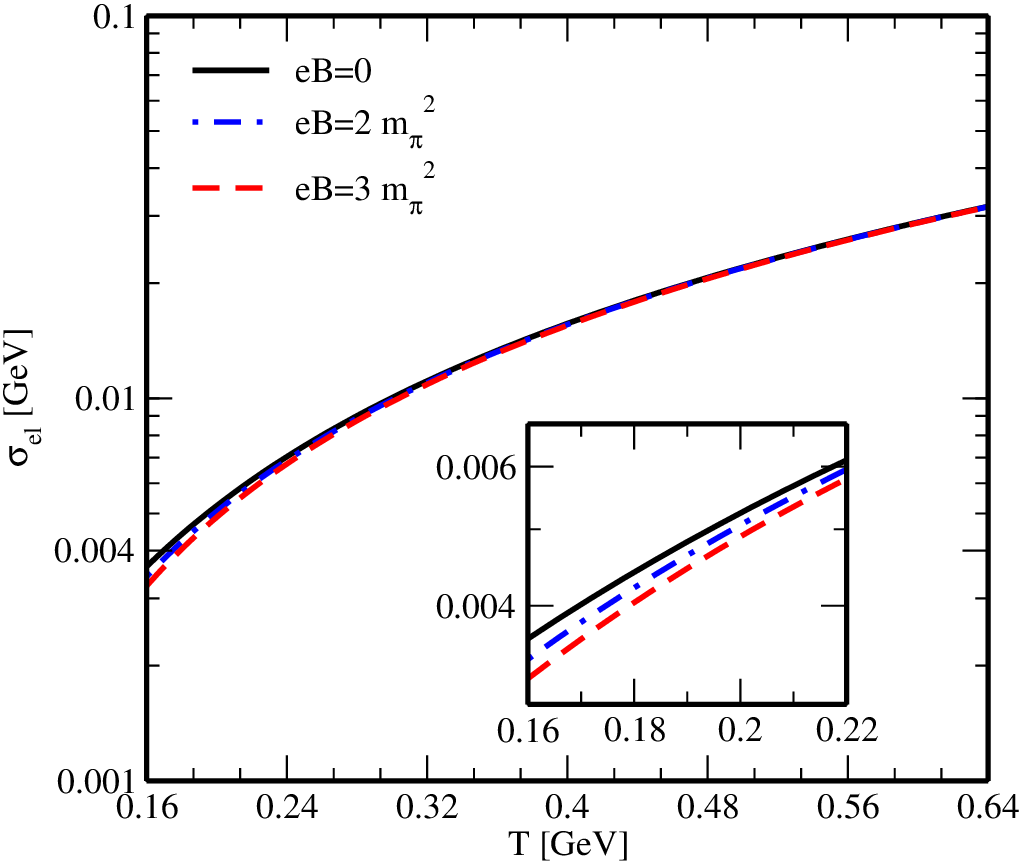}&
\hspace{0.74 cm}
\includegraphics[width=6.86cm]{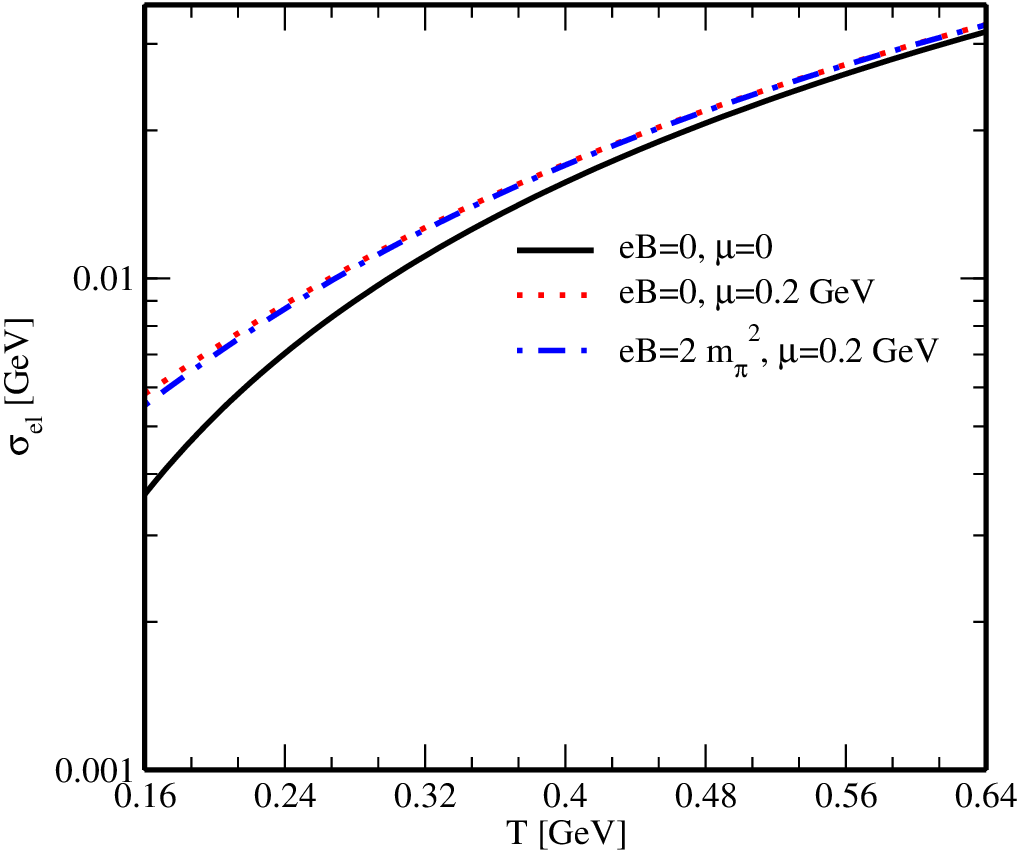} \\
a & b
\end{tabular}
\caption{The variation of the electrical conductivity, $\sigma_{\rm el}$ with temperature 
(a) in the presence of weak magnetic field and (b) in the presence 
of finite chemical potential.}\label{Fig.1}
\end{center}
\end{figure}

\begin{figure}[]
\begin{center}
\begin{tabular}{c c}
\includegraphics[width=6.86cm]{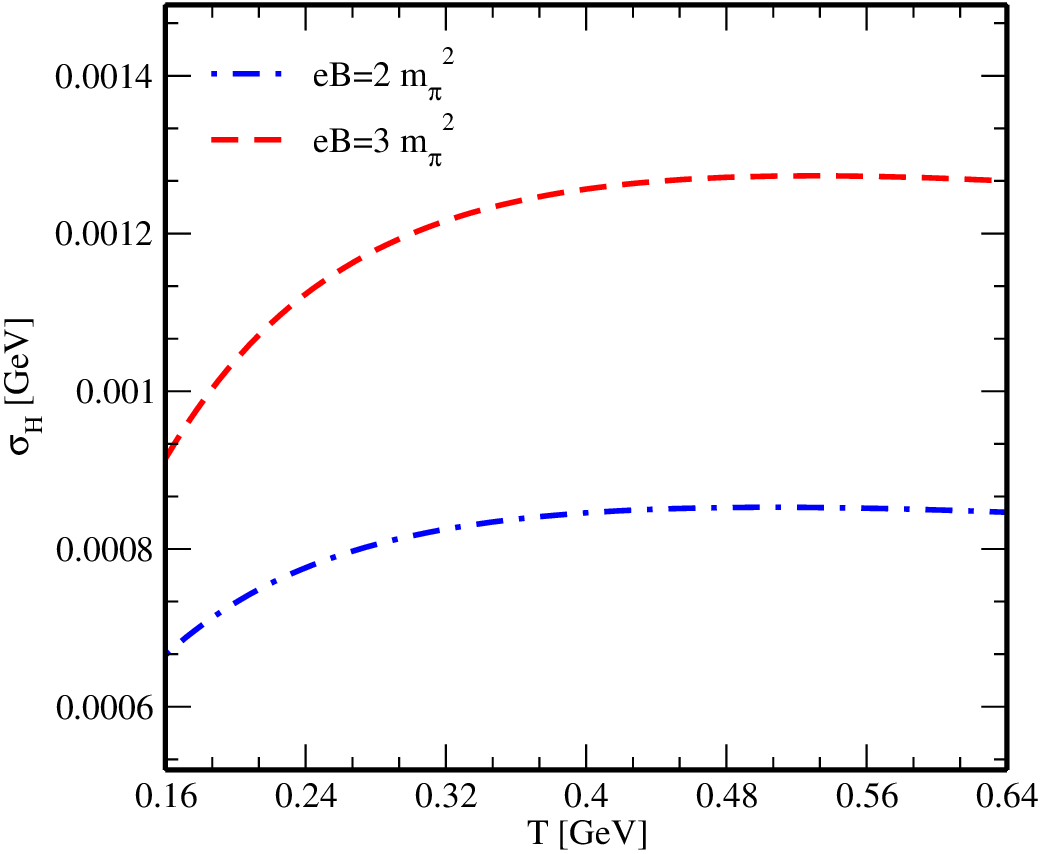}&
\hspace{0.74 cm}
\includegraphics[width=6.86cm]{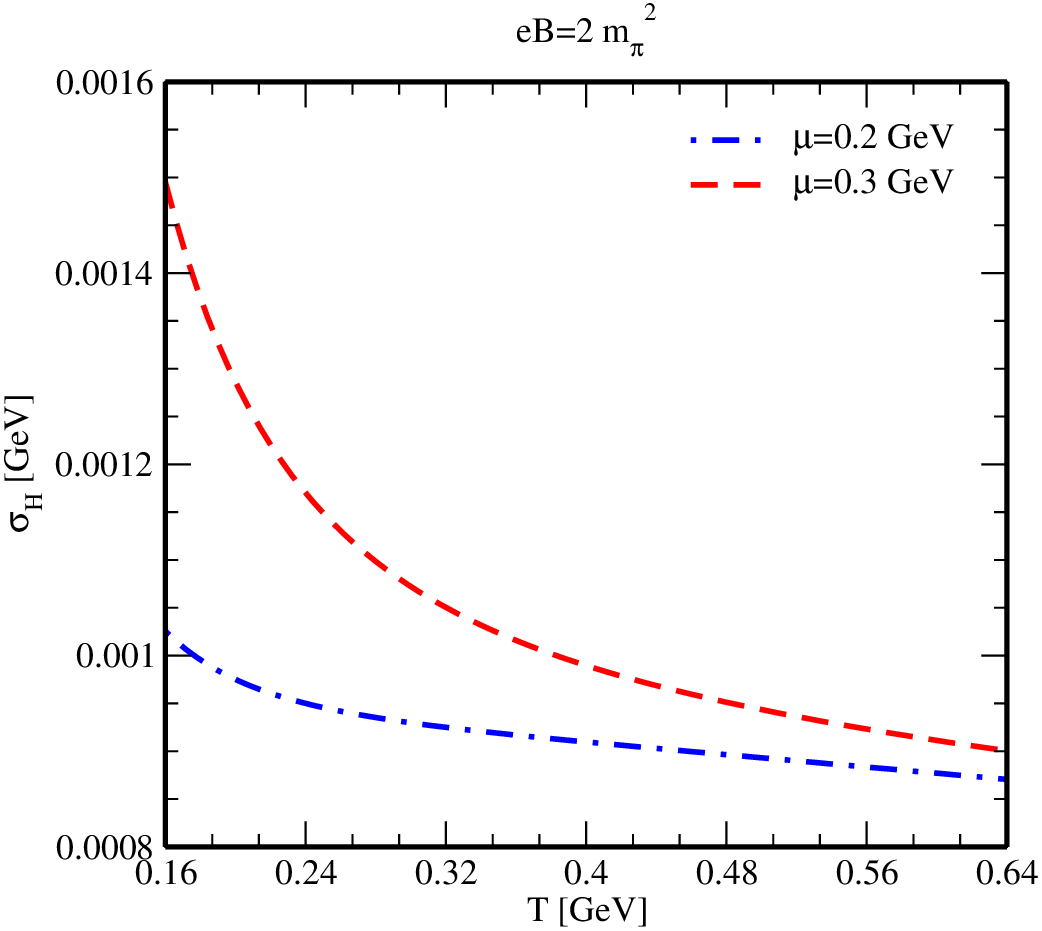} \\
a & b
\end{tabular}
\caption{The variation of the Hall conductivity, $\sigma_{\rm H}$ with temperature 
(a) in the presence of weak magnetic field and (b) in the presence 
of finite chemical potential.}\label{Fig.11}
\end{center}
\end{figure}

Figure \ref{Fig.1} shows the variation of the electrical conductivity, $\ec$ as a 
function of the temperature in the presence of a weak magnetic field. In 
particular, figure \ref{Fig.1}a shows the variation of $\ec$ at zero 
chemical potential while figure \ref{Fig.1}b shows the same for finite chemical 
potential. It can be observed from figure \ref{Fig.1}a that the $\ec$
slightly decreases in the presence of a weak magnetic field when compared to  
zero magnetic field case at low temperatures. This observation can be related to the motion of electrically charged quarks in the hot QCD medium.
According to the Ohm's law, the electrical conductivity is directly proportional to the current along the 
direction of the electric field. But in the presence of a magnetic 
field, quarks experience a Lorentz force, which influences their direction of motion 
(of quarks) and results in a reduction of electric current in the direction of electric field, and hence a decrease in the electrical conductivity is observed. Figure \ref{Fig.1}b shows that the presence of a finite 
chemical potential results in the increase of $\ec$ in a weakly magnetized 
QCD medium. Thus, it is inferred that the weak magnetic field decreases the 
charge conduction in a hot QCD matter, whereas the finite chemical potential 
tends to increase it. This can be comprehended from the fact that, 
in the weak magnetic field regime, the energy scale associated with the 
magnetic field is smaller than other energy scales, so $\ec$ becomes more 
sensitive to the energy scale related to the chemical potential and it 
results in the increase of the charge transport at finite chemical 
potential even in the presence of a weak magnetic field. The behavior of 
$\ec$ is also modulated by the distribution function, because, according to 
the nonrelativistic Drude's formula, the electrical conductivity is 
directly proportional to the number density, {\em i.e.} the integration of 
distribution function over momentum space. Thus, the decrease of 
$\sigma_{\rm el}$ due to weak magnetic field and its increase due to 
finite chemical potential can also be understood from the decrease of 
distribution function in a weak magnetic field and its increase at finite 
chemical potential (figures \ref{dftu} and \ref{dfpu}). 

In figure \ref{Fig.11}, the variation of Hall conductivity, $\eh$ as a 
function of temperature is shown in the presence of weak magnetic field and 
finite chemical potential. It can be observed that, at $\mu=0$, $\eh$ shows increasing behavior with $T$ in the low temperature regime (figure \ref{Fig.11}a), which is mainly due to the increase of distribution function with $T$. On the other hand, $\eh$ shows decreasing behavior with $T$ at finite $\mu$ (figure \ref{Fig.11}b), which can be comprehended as follows. At finite $\mu$, the increase of distribution function with $T$ becomes smaller and the factor $\beta\tau_f^2$ ($\approx\frac{1}{T^3}$, at least in the weak magnetic field regime) appearing in the expression of $\eh$ \eqref{H.Conductivity} dominates, thus an overall decrease of $\eh$ with increasing $T$ is observed. 

Hall conductivity vanishes at zero magnetic field, which can be understood 
from the presence of cyclotron frequency ($\omega_c$) in the numerator 
of eq. \eqref{H.Conductivity}. With the increase of magnetic field, $\eh$ gets 
increased even in the weak magnetic field limit (figure \ref{Fig.11}a), unlike the case of $\ec$, because $\eh$ is directly related to $\omega_c$. Increase in the magnitude of $\eh$ is also observed with the increase of chemical potential (figure \ref{Fig.11}b), which can be inferred from the fact that, at finite chemical potential, difference in the numbers of particles and antiparticles produces a net Hall current which is proportional to the Hall conductivity. 

\subsection{Components of heat transport}
\begin{figure}[]
\begin{center}
\begin{tabular}{c c}
\includegraphics[width=6.86cm]{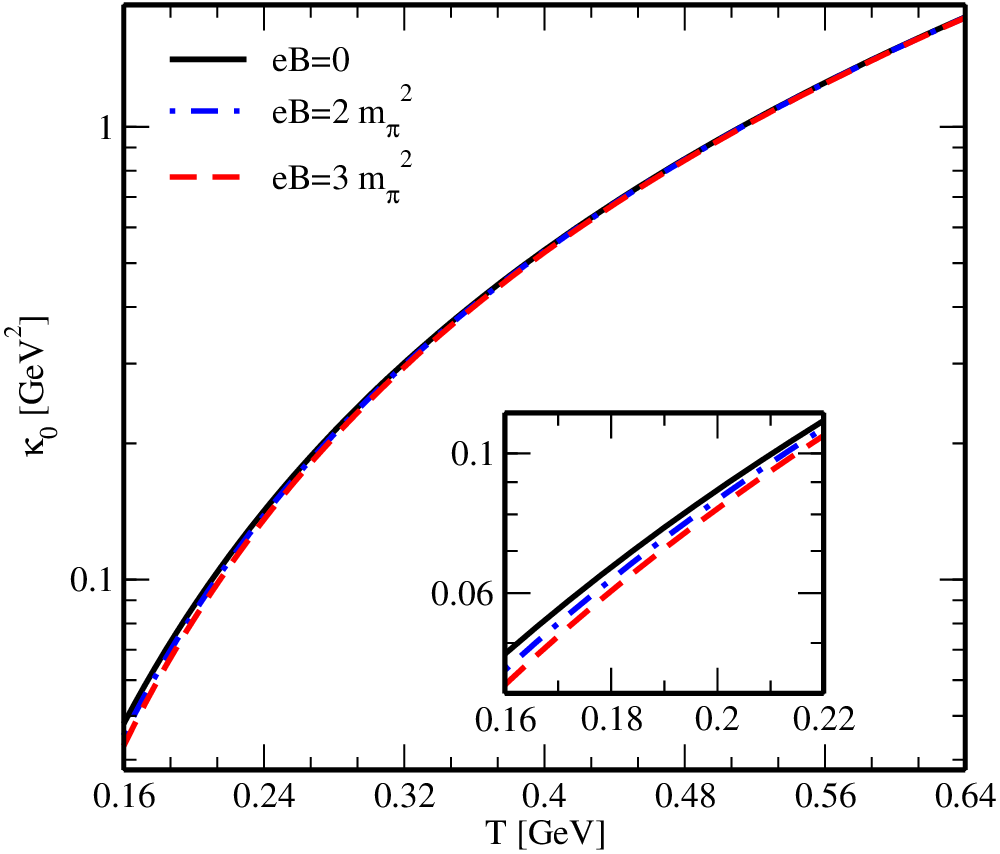}&
\hspace{0.74 cm}
\includegraphics[width=6.86cm]{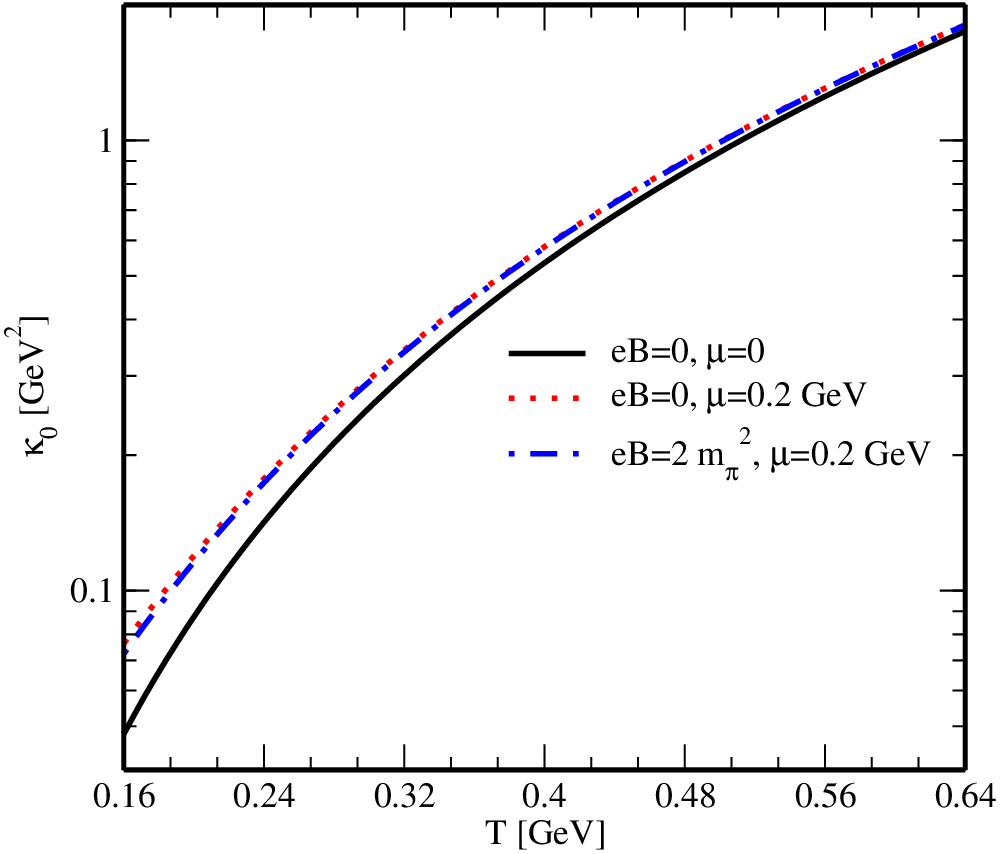} \\
a & b
\end{tabular}
\caption{The variation of the thermal conductivity, $\kappa_0$ 
with temperature (a) in the presence of weak magnetic field 
and (b) in the presence of finite chemical potential.}\label{Fig.2}
\end{center}
\end{figure}

\begin{figure}[]
\begin{center}
\begin{tabular}{c c}
\includegraphics[width=6.86cm]{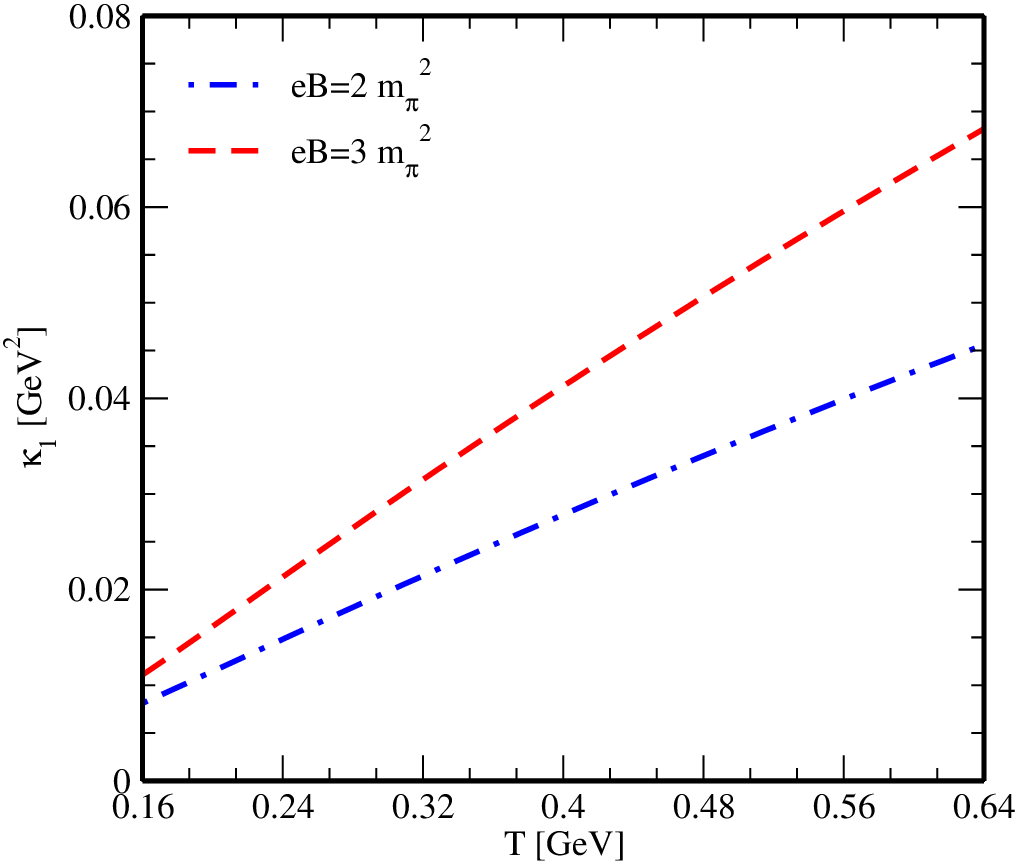}&
\hspace{0.74 cm}
\includegraphics[width=6.86cm]{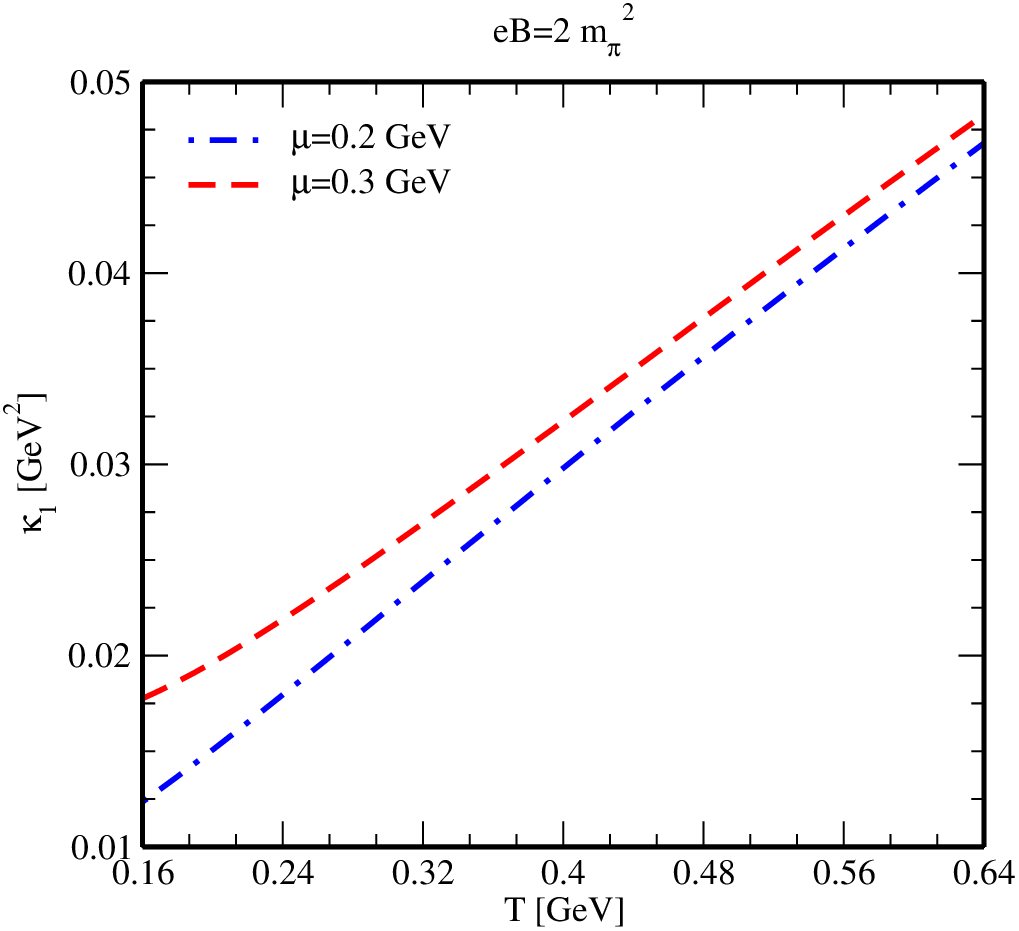} \\
a & b
\end{tabular}
\caption{The variation of the Hall-type thermal conductivity, $\kappa_1$ 
with temperature (a) in the presence of weak magnetic field and (b) in 
the presence of finite chemical potential.}\label{Fig.22}
\end{center}
\end{figure}

Figure \ref{Fig.2} depicts the variation of the thermal conductivity, $\kappa_0$ 
with the temperature in the presence of a weak magnetic field. 
It can be observed that the presence of weak magnetic field reduces 
$\kappa_0$ and the reduction is larger at low temperatures 
(figure \ref{Fig.2}a). On the other hand, a finite chemical potential 
enhances the magnitude of $\kappa_0$ (figure \ref{Fig.2}b). However, this 
enhancement of $\kappa_0$ is not same over the entire range of temperature, 
rather it is more conspicuous at low temperatures. In figure \ref{Fig.22}, the 
Hall-type thermal conductivity, $\kappa_1$ is plotted as a function of temperature 
and it vanishes at zero magnetic field, which can be understood from 
eq. \eqref{H.C.(1)}. Unlike $\kappa_0$, $\kappa_1$ directly depends on 
cyclotron frequency, so one can observe that the 
magnitude of $\kappa_1$ gets increased with the magnetic field even in the weak 
magnetic field regime (figure \ref{Fig.22}a). In addition, finite chemical 
potential also increases the magnitude of $\kappa_1$ (figure \ref{Fig.22}b). Both $\kappa_0$ and $\kappa_1$ are found to increase with $T$, but one would expect the reverse effect of $T$ on these conductivities due to the presence of terms $\beta^2\tau_f$ and $\beta^2\tau_f^2$ in the expressions of $\kappa_0$ \eqref{H.C.} and $\kappa_1$ \eqref{H.C.(1)}, respectively. However, in their expressions, the increase of enthalpy per particle and the increase of distribution function with $T$ somewhat compensate this effect and thus, we observe an overall increase of $\kappa_0$ and $\kappa_1$ with temperature. 

In the strong magnetic field regime, there is a severe reduction of the phase space from (3+1)-dimensions to (1+1)-dimensions, so the charged particles are constrained to move along the direction of magnetic field. On the other hand, the weak magnetic field does not restrict the 3-dimensional dynamics and there exist different components of charge and heat transport coefficients. Thus, it may not be plausible to compare our results on conductivities in the weak magnetic field regime with that in the strong magnetic field regime at the equal base. However, we may roughly compare our results in the weak magnetic field with the results obtained in the strong magnetic field. For example, in ref. \cite{Buividovich:PRL105'2010}, $\ec$ was calculated using the quenched SU($2$) lattice gauge theory, where for the deconfinement phase, only the component of $\ec$ along the direction of magnetic field exists and it increases as the strength of magnetic field increases. According to the observation on $\ec$ which was calculated in ref. \cite{Nam:PRD86'2012} using the Kubo formalism and the dilute instanton-liquid model, the effect of the external magnetic field was relatively considerable in the low temperature region $T\leq200$ MeV. In ref. \cite{Hattori:PRD94'2016}, $\ec$ was estimated using the real time formalism through the diagrammatic method and found to be much larger than its counterpart at zero magnetic field. The main reasons behind this large increment were the strong magnetic field and the smaller value of the current quark mass. In ref. \cite{Fukushima:PRL120'2018}, the Landau level resummation via kinetic equations has been implemented and a fixed QCD coupling ($\alpha_s=0.3$) was used in the evaluation of $\ec$, which has very large value in the lowest Landau level (LLL) approximation and remains almost insensitive to $\mu$, unlike our observation in weak magnetic field, where the effect of $\mu$ is noticeable. In ref. \cite{Rath:PRD100'2019}, electrical and thermal conductivities have been studied using the kinetic theory with the quasiparticle model in the strong magnetic field limit, where both the conductivities were found to be enhanced by the strong magnetic field, unlike our present work, where they get decreased in weak magnetic field as compared to their counterparts in the absence of magnetic field. The effective fugacity approach has also reported large magnitudes of $\ec$ ($\mathcal{O}(10^2)$) \cite{Kurian:PRD99'2019} and $\kappa$ (between $\mathcal{O}(10^1)$ to $\mathcal{O}(10^2)$) \cite{Kurian:EPJC79'2019} in the LLL approximation due to the presence of strong magnetic field. Our results in the weak magnetic field regime are totally different from the abovementioned results in the strong magnetic field regime. Main reasons behind this are the differences in relaxation times, phase spaces and distribution functions in both types of magnetic field regimes. 

\section{Applications}
In this section, we study the effects of weak magnetic field and chemical 
potential on some applications of charge and heat transport coefficients. 
Subsections 5.1, 5.2 and 5.3 are devoted to observe the local equilibrium 
property of the medium through the Knudsen number, the elliptic flow and 
the relative behavior between the charge conduction and the heat 
conduction through the Wiedemann-Franz law, respectively. 

\subsection{Knudsen number}
The Knudsen number, $\Omega$ is defined in terms of the mean free path ($\lambda$) 
and the characteristic length scale of the medium ($l$) as
\begin{eqnarray}
\Omega=\frac{\lambda}{l}
~.\end{eqnarray}
If the mean free path is smaller than the characteristic length scale, then $\Omega$ is 
less than one and the equilibrium hydrodynamics is applicable. The mean free path 
can be calculated using $\kappa$ as
\begin{eqnarray}\label{Dependence}
\lambda=\frac{3\kappa}{vC_V}
~,\end{eqnarray}
where $C_V$ and $v$ represent the specific heat at constant volume 
and the relative speed, respectively. Thus, $\Omega$ takes the 
following form, 
\be
\Omega=\frac{3\kappa}{lvC_V}
~.\ee
Here, $C_V$ has been calculated from the energy-momentum tensor 
($C_V=\partial (u_\mu T^{\mu\nu}u_\nu)/\partial T$). In 
computing $\Omega$, we have fixed $v\simeq 1$ and $l=4$ fm. 
We note that, for two heat transport coefficients ($\kappa_0$ and $\kappa_1$) 
in a weak magnetic field, there also exist two components of the Knudsen number, 
such as $\Omega_0$ and $\Omega_1$. 

\begin{figure}[]
\begin{center}
\begin{tabular}{c c}
\includegraphics[width=6.86cm]{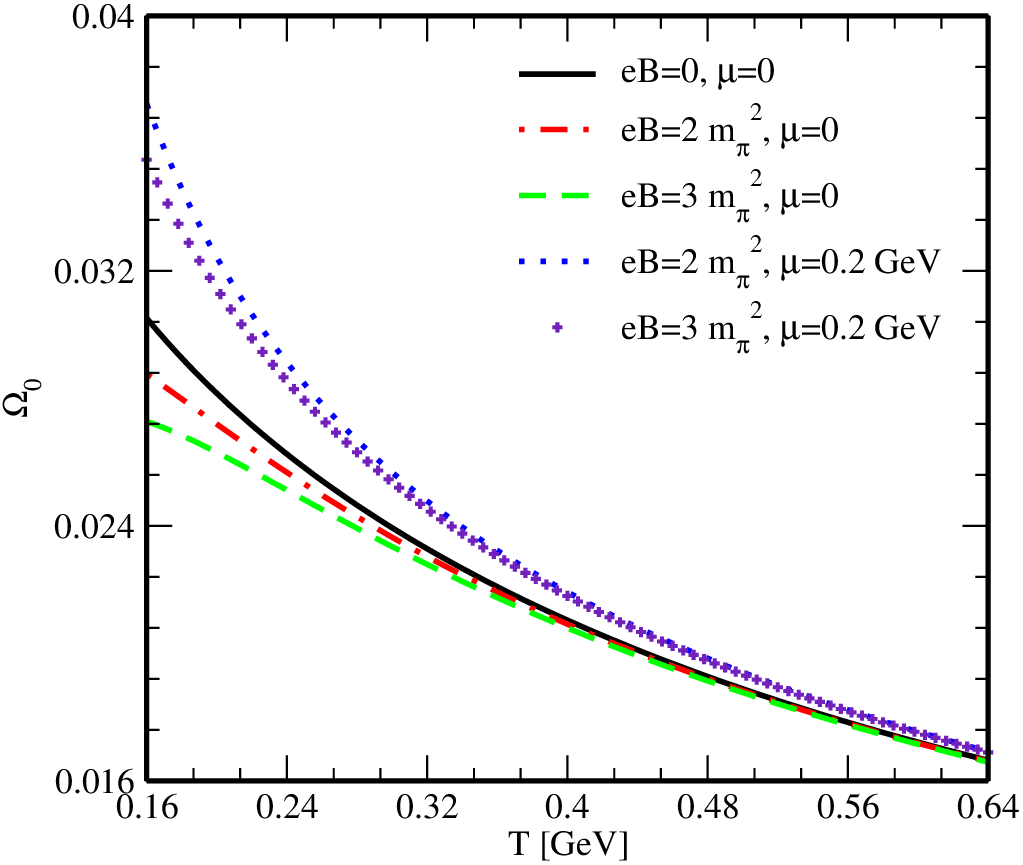}&
\hspace{0.74 cm}
\includegraphics[width=6.86cm]{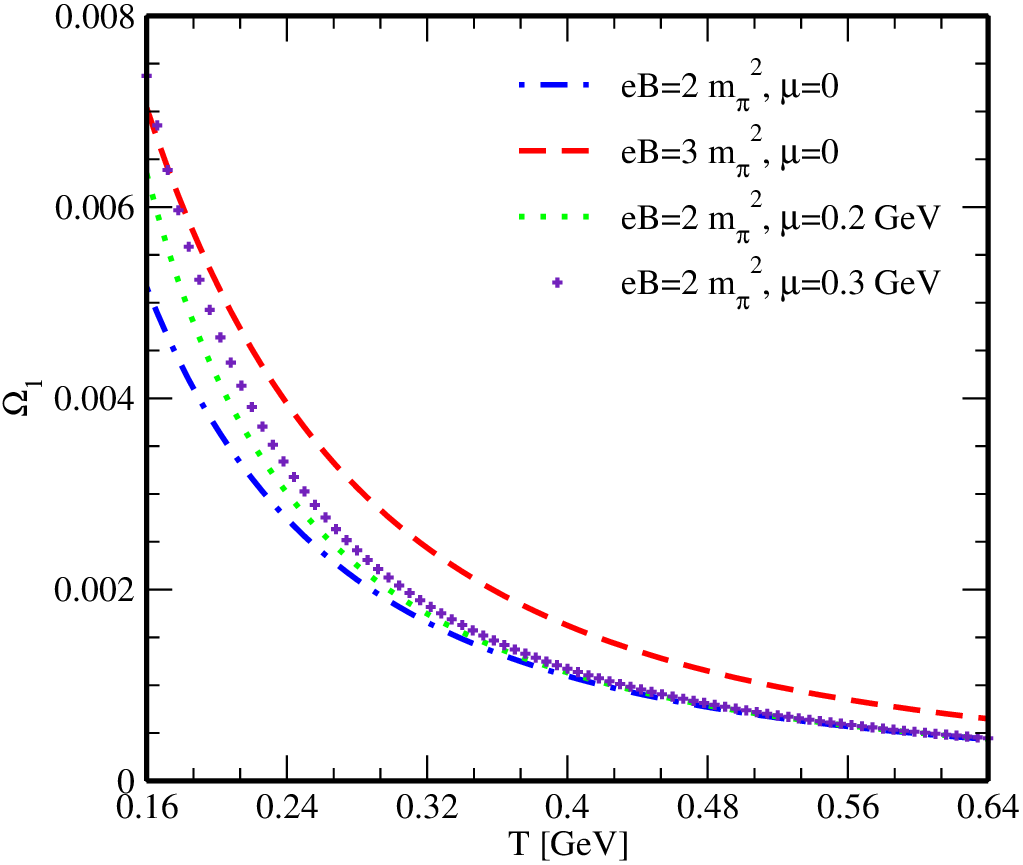} \\
a & b
\end{tabular}
\caption{The variations of (a) $\Omega_0$ and (b) $\Omega_1$ with temperature 
for different values of magnetic field and chemical potential.}\label{Fig.3}
\end{center}
\end{figure}

Figure \ref{Fig.3} shows the variations of the Knudsen number components $\Omega_0$ 
and $\Omega_1$ as functions of temperature. It can be seen that $\Omega_0$ lies 
much below unity in the absence of both magnetic field and chemical potential 
(figure \ref{Fig.3}a). The presence of weak magnetic field decreases $\Omega_0$, 
whereas the additional presence of chemical potential leads to an increase 
of this component of the Knudsen number. Similarly, $\Omega_1$ lies much 
below unity and is even smaller in magnitude as compared to $\Omega_0$. 
Like $\Omega_0$, $\Omega_1$ also increases with the rise of chemical potential. The observations on $\Omega_0$ and $\Omega_1$ at finite $eB$ and finite $\mu$ corroborate the observations on $\kappa_0$ (figure \ref{Fig.2}) and $\kappa_1$ (figure \ref{Fig.22}) in the similar environment. Both, $\Omega_0$ and $\Omega_1$ remain much less than unity over the entire range of temperature. It indicates that, there is sufficient separation between microscopic and macroscopic length scales of the medium and the hot QCD matter stays in local equilibrium even in the presence of both weak magnetic field and finite chemical potential. In all cases, $\Omega_0$ and $\Omega_1$ follow the decreasing trend with the rise 
of temperature, same as in the absence of magnetic field. 

\subsection{Elliptic flow}
The elliptic flow coefficient, $v_{2}$ describes the azimuthal anisotropy in the 
momentum space of the produced particles in heavy ion collisions. This is a 
direct consequence of initial pressure gradients because of the initial spatial 
anisotropy and the interactions among the produced particles 
\cite{Ollitrault:PRD46'1992}. The elliptic flow is related to the Knudsen number 
by the following expression \cite{Bhalerao:PLB627'2005,Drescher:PRC76'2007,Gombeaud:PRC77'2008}, 
\be\label{Elliptic flow}
v_2=\frac{v_2^h}{1+\frac{\Omega}{\Omega_h}}
,\ee
where $v_2^h$ is the value of elliptic flow in the hydrodynamic limit, 
{\em i.e.} $\Omega\rightarrow 0$ limit. We note that we have 
used $\Omega=\Omega_0$ here. The value of $\Omega_h$ can be 
obtained by observing the transition between the hydrodynamic regime 
and the free streaming particle regime. The presence of weak magnetic 
field as well as chemical potential could significantly affect the 
magnitude of elliptic flow. In our calculation, we have used $v_2^h \approx 0.1$ and $\Omega_h \approx 0.7$, which are obtained from the transport calculation in ref. \cite{Gombeaud:PRC77'2008}. 

\begin{figure}[]
\begin{center}
\begin{tabular}{c c}
\includegraphics[width=6.86cm]{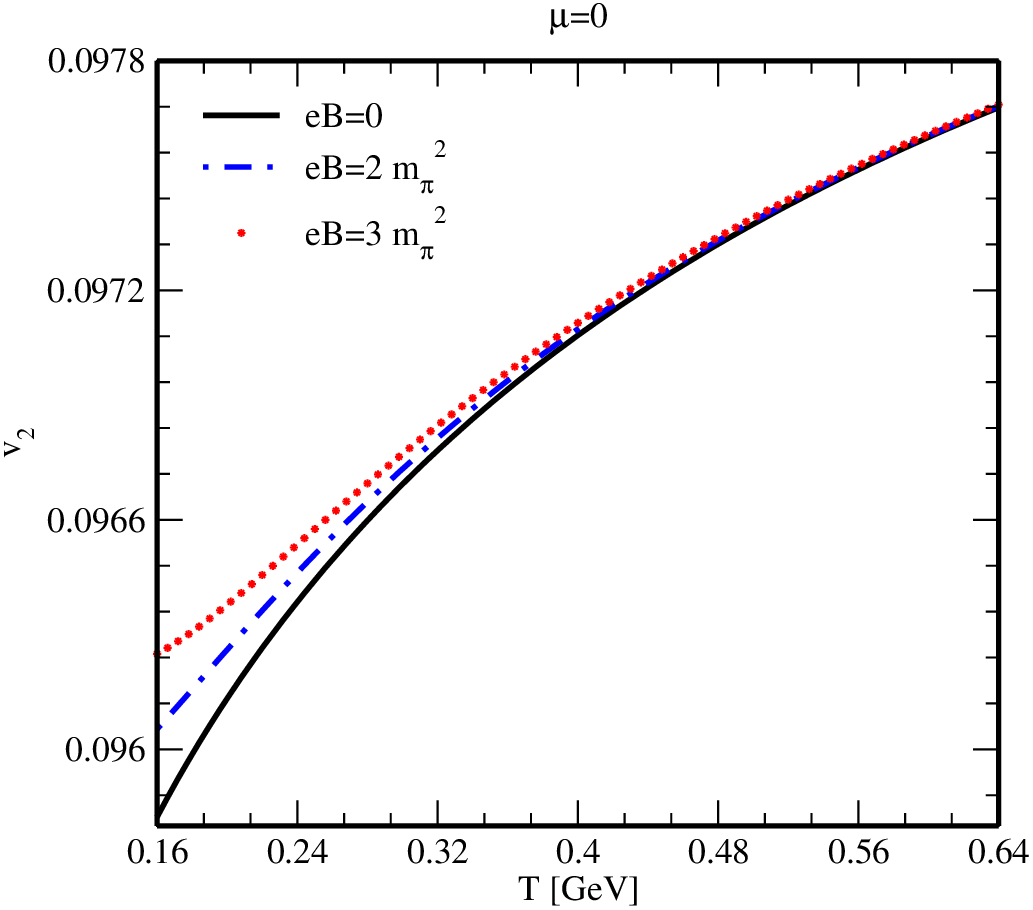}&
\hspace{0.74 cm}
\includegraphics[width=6.86cm]{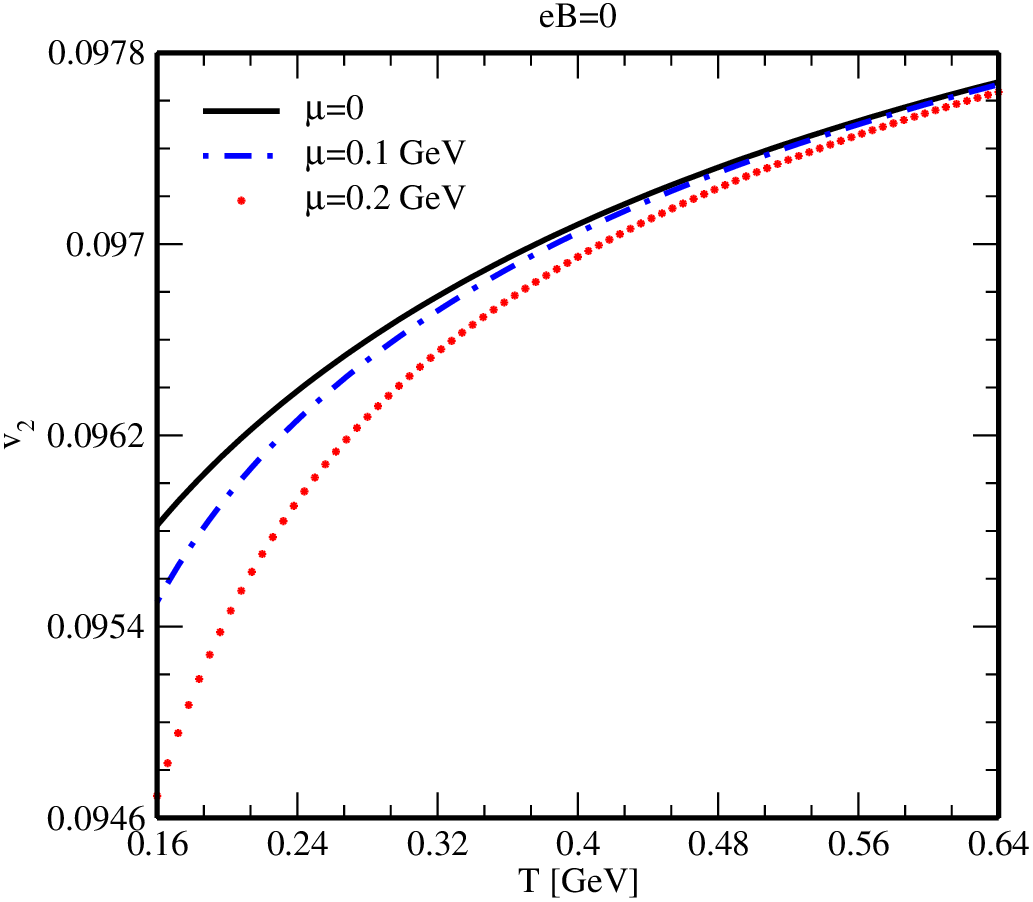} \\
a & b
\end{tabular}
\caption{The elliptic flow coefficient, $v_2$ as a function of temperature 
(a) in the presence of weak magnetic field and (b) at 
finite chemical potential.}\label{E.F.}
\end{center}
\end{figure}

Figure \ref{E.F.} depicts the variation of elliptic flow as a function 
of temperature for weak magnetic field and finite chemical potential. 
It can be observed that $v_2$ increases with temperature. 
Figure \ref{E.F.}a compares the temperature dependence of $v_2$ for different values of magnetic field. It can be seen that 
$v_2$ is higher for weak magnetic field at low temperatures. The increase 
is not uniform over the entire range of temperature. The $v_2$  approaches its value at zero magnetic field at higher temperatures, because the sensitivity of $v_2$ to the weak magnetic field decreases as the energy scale related to the temperature grows. The presence of the strong magnetic field also enhances the elliptic flow, as it has been reported in 
ref. \cite{Mohapatra:MPLA26'2011}. The effect of  finite chemical potential on the elliptic flow is displayed in figure \ref{E.F.}b. The finite chemical potential  decreases the magnitude of elliptic flow. The deviation of $v_2$ from its value at zero chemical potential is larger at lower temperatures and the convergence happens at higher temperatures. 

As observed in figure \ref{Fig.3}, the Knudsen number becomes zero when the 
hydrodynamic limit is approached. In this case, $v_{2}$ becomes maximum and 
attains its hydrodynamic value. It can be inferred, that, as $\Omega$ 
decreases, the number of collisions increases which leads to 
a larger anisotropic flow, hence $v_2$ grows and eventually saturates when the 
medium reaches local equilibrium. One can also understand from 
eq. \eqref{Elliptic flow} that, for large $\Omega$ or far from the hydrodynamic 
limit, $v_2$ decreases like $1/\Omega$. The increase in $v_{2}$ at 
finite magnetic field can be comprehended as follows: The presence of magnetic 
field makes significant variations in the velocities of particles and the degree 
of variation depends on the angle between the direction of flow and the direction of 
magnetic field which impacts  the development of anisotropy, thus, an increase in the 
elliptic flow is observed. To some extent, this 
enhancement of $v_2$ in the weak magnetic field regime can also be understood 
from the reduction of thermal conductivity (through the dependence on mean free 
path in eq. \eqref{Dependence}) in the similar environment (figure \ref{Fig.2}a). 
However, with an increase of chemical potential, thermal conductivity increases 
(figure \ref{Fig.2}b), which results in the decrease of $v_2$ at finite 
chemical potential. Thus, the magnitude of $v_2$ acts as a probe to detect the 
level of thermalization in different scenarios. Quantitatively, for a temperature 
range of 0.16-0.64 GeV, our calculation gives the value of $v_2$ in the 
range 0.0958-0.0976 at $eB=0$, $\mu=0$, in the range 0.0946-0.0976 at $eB=0$, 
$\mu\neq 0$ and in the range 0.0962-0.0976 at $eB\neq 0$, $\mu=0$. The obtained 
values are closer to the experimental data of elliptic flow obtained from STAR 
collaboration at RHIC \cite{Snellings:APHA21'2004,Tang:0808.2144} and 
ALICE collaboration at LHC \cite{Aamodt:PRL105'2010} for $p_T\approx1$ GeV. 

\subsection{Wiedemann-Franz law}
According to the Wiedemann-Franz law, the ratio of the heat transport coefficient to the 
charge transport coefficient is directly proportional to the temperature, where the 
proportionality factor is the Lorenz number ($L$), 
\begin{eqnarray}
\frac{\kappa}{\sigma_{\rm el}}=LT
~.\end{eqnarray}
This law sheds light on the relative behavior of the charge transport and the 
heat transport in a medium. We note that, in weak magnetic field, there 
exist two components of the Lorenz number, such as $L_0$ and $L_1$, 
\be
&&L_0=\frac{\kappa_0}{\sigma_{\rm el}T} ~, \\
&&L_1=\frac{\kappa_1}{\sigma_{\rm H}T}
~.\ee

\begin{figure}[]
\begin{center}
\begin{tabular}{c c}
\includegraphics[width=6.86cm]{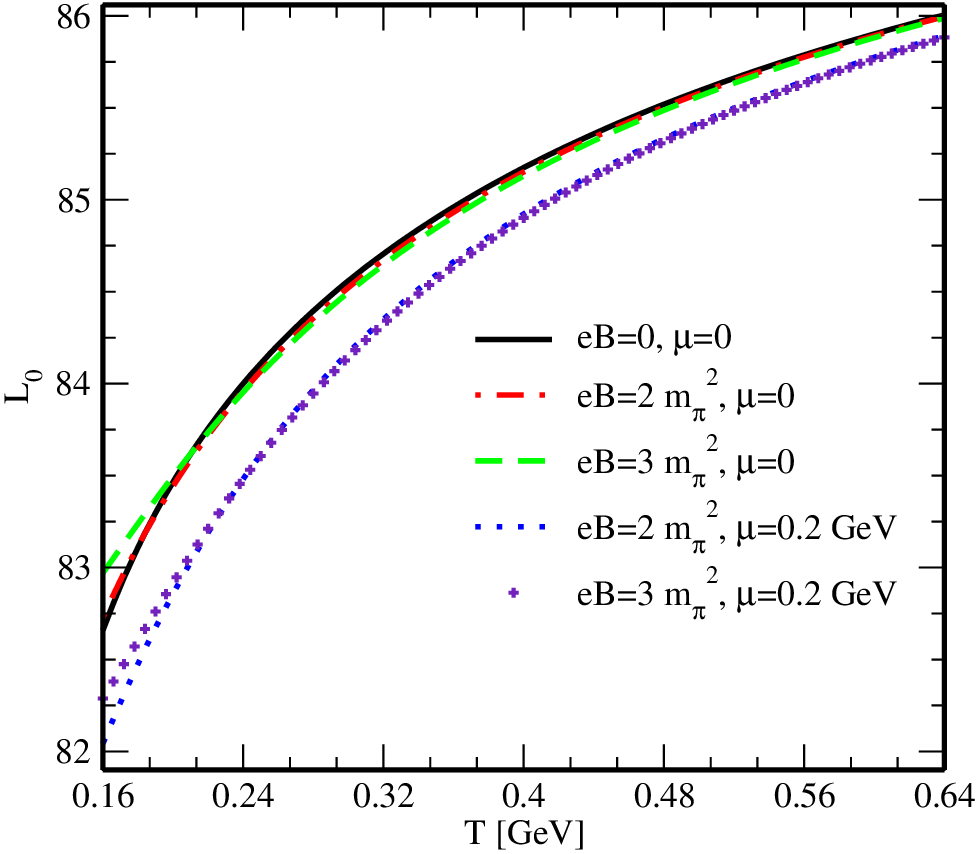}&
\hspace{0.74 cm}
\includegraphics[width=6.86cm]{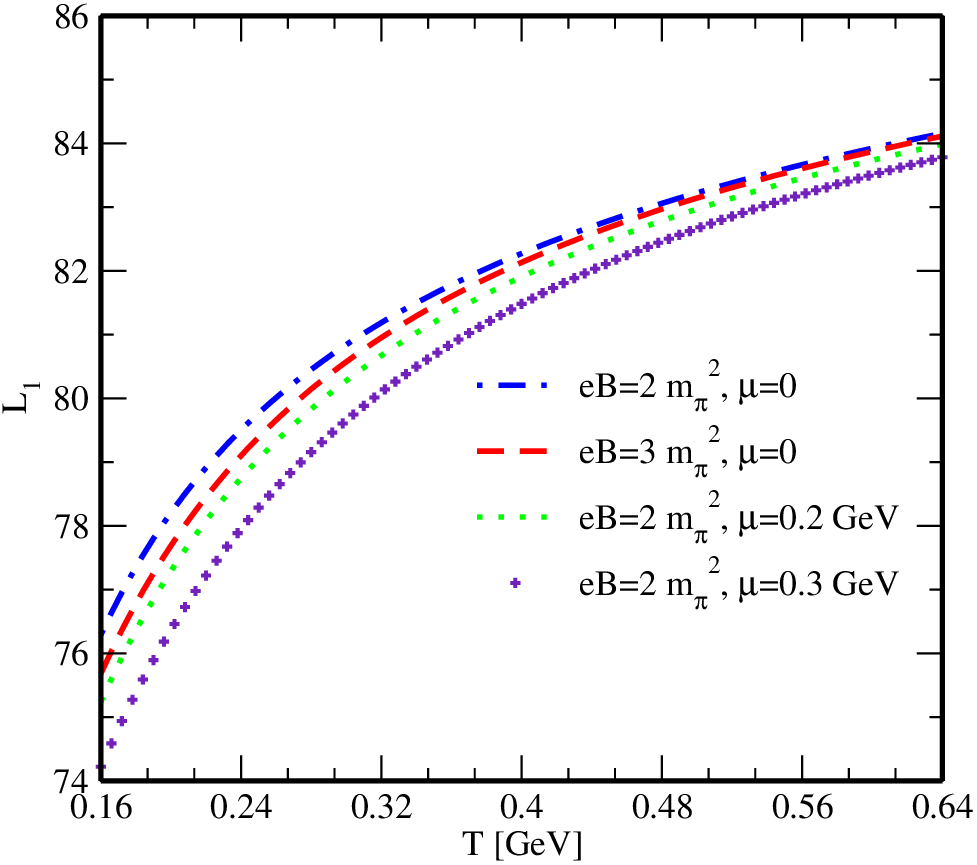} \\
a & b
\end{tabular}
\caption{The variations of (a) $L_0$ and (b) $L_1$ with temperature 
for different values of magnetic field and chemical potential.}\label{Fig.4}
\end{center}
\end{figure}

Figure \ref{Fig.4} depicts the variations of the Lorenz number components $L_0$ 
and $L_1$ as functions of the temperature for various conditions of magnetic 
field and chemical potential. It can be observed that both $L_0$ and $L_1$ 
increase with $T$, {\em i.e.} the heat transport forges ahead of the 
charge transport. This indicates that the hot QCD matter does not comply 
with the Wiedemann-Franz law. Both $L_0$ and $L_1$ of the weakly magnetized 
medium almost follow the same trend as that of the zero magnetic field case 
with temperature. The presence of finite chemical potential in the magnetized 
medium reduces the magnitudes of $L_0$ and $L_1$ for all temperatures. 
However, throughout the variation, these components of the Lorenz number remain 
larger than unity, indicating that the thermal conductivity ($\kappa_0$) 
prevails over the electrical conductivity ($\ec$) and the Hall-type thermal 
conductivity ($\kappa_1$) also prevails over the Hall conductivity ($\eh$) at any 
value of the temperature. The observed enhanced difference between the 
heat transport and the charge transport is in accordance with the observations on 
charge conduction behavior in figures \ref{Fig.1} and \ref{Fig.11} and heat 
conduction behavior in figures \ref{Fig.2} and \ref{Fig.22}. 

\section{Summary}
The effects of weak magnetic field and finite chemical potential on the charge and heat transport properties of hot and dense QCD matter have been investigated. In the presence of magnetic field, the transport coefficients do not remain isotropic and they possess different components. The components of charge transport and heat transport, such as electrical conductivity ($\ec$), Hall conductivity ($\eh$), thermal conductivity ($\kappa_0$) and Hall-type thermal conductivity ($\kappa_1$) were determined by solving the relativistic Boltzmann transport equation within the relaxation time approximation of kinetic theory. A reduction in the magnitudes of $\ec$ and $\kappa_0$ and an enhancement in the magnitudes of $\eh$ and $\kappa_1$ with the increase of magnetic field were observed, whereas the emergence of finite chemical potential tends to increase all of their magnitudes. The transport coefficients were further used to study the Knudsen number, the elliptic flow and the Wiedemann-Franz law. The Knudsen number components in the weakly magnetized hot and dense QCD matter retain their values much below unity. Thus, there is sufficient separation between the mean free path and the characteristic length scale for the medium to remain in the local equilibrium state. The elliptic flow gets increased in the presence of the weak magnetic field, whereas the presence of finite chemical potential decreases it. The Lorenz number components in the Wiedemann-Franz law were found to be strongly affected by the chemical potential than by the weak magnetic field. However, with the increase of temperature, the Lorenz number components were observed to increase, confirming the violation of the Wiedemann-Franz law for hot and dense QCD matter in the presence of a weak magnetic field. 

\section{Acknowledgment}
One of us (S. R.) would like to acknowledge the Indian Institute of Technology Bombay 
for the Institute postdoctoral fellowship. 

\appendix
\appendixpage
\addappheadtotoc
\begin{appendices}
\renewcommand{\theequation}{A.\arabic{equation}}
\section{Derivation of equation \eqref{deltaf.q}}\label{I.C. of Q.D.F.1}
By substituting the following partial derivatives in eq. \eqref{R.B.T.E.(4)}, 
\be
\nonumber v_x\frac{\partial f_f}{\partial p_0} &=& -\beta v_xf_f^0-qE\tau_f\beta f_f^0v_x^2\left(\frac{1}{\omega_f}+\beta\right)-\beta f_f^0\Gamma_xv_x^2\left(\frac{1}{\omega_f}+\beta\right) \\ && -\beta f_f^0\Gamma_yv_xv_y\left(\frac{1}{\omega_f}+\beta\right)-\beta f_f^0\Gamma_zv_xv_z\left(\frac{1}{\omega_f}+\beta\right), \\
\nonumber v_x\frac{\partial f_f}{\partial p_y} &=& -\beta v_xv_yf_f^0-qE\tau_f\beta f_f^0v_x^2v_y\left(\frac{1}{\omega_f}+\beta\right)-\beta f_f^0\Gamma_xv_x^2v_y\left(\frac{1}{\omega_f}+\beta\right) \\ && -\beta f_f^0\Gamma_yv_xv_y^2\left(\frac{1}{\omega_f}+\beta\right)+\frac{v_x\Gamma_y\beta f_f^0}{\omega_f}-\beta f_f^0\Gamma_zv_xv_yv_z\left(\frac{1}{\omega_f}+\beta\right) , \\
\nonumber v_y\frac{\partial f_f}{\partial p_x} &=& -\beta v_yv_xf_f^0-qE\tau_f\beta f_f^0v_yv_x^2\left(\frac{1}{\omega_f}+\beta\right)+\frac{qE\tau_f\beta f_f^0v_y}{\omega_f} \\ && \nonumber -\beta f_f^0\Gamma_xv_yv_x^2\left(\frac{1}{\omega_f}+\beta\right)+\frac{\Gamma_x\beta f_f^0v_y}{\omega_f}-\beta f_f^0\Gamma_yv_y^2v_x\left(\frac{1}{\omega_f}+\beta\right) \\ && -\beta f_f^0\Gamma_zv_yv_zv_x\left(\frac{1}{\omega_f}+\beta\right)
,\ee
and then dropping higher order velocity terms, we get
\be\label{R.B.T.E.(5)}
-qE\tau_fv_x+\left(\Gamma_xv_x+\Gamma_yv_y+\Gamma_zv_z\right)-\frac{qB\tau_f}{\omega_f}\left(v_x\Gamma_y-v_y\Gamma_x\right)+\frac{\tau_f^2qBqEv_y}{\omega_f}=0
.\ee
Comparing the coefficients of $v_z$ on both sides of eq. \eqref{R.B.T.E.(5)}, we get $\Gamma_z=0$. Then, we have
\be\label{R.B.T.E.(6)}
-qEv_x+\frac{\Gamma_x}{\tau_f}v_x-\omega_c\Gamma_yv_x+\frac{\Gamma_y}{\tau_f}v_y+\omega_c\Gamma_xv_y+\tau_f\omega_cqEv_y=0
,\ee
where the cyclotron frequency, $\omega_c=\frac{qB}{\omega_f}$. Equating 
coefficients of $v_x$ and $v_y$ on both sides of 
eq. \eqref{R.B.T.E.(6)}, we get
\be
&&\label{vx}\frac{\Gamma_x}{\tau_f}-\omega_c\Gamma_y-qE=0, \\ 
&&\label{vy}\frac{\Gamma_y}{\tau_f}+\omega_c\Gamma_x+\tau_f\omega_cqE=0
.\ee
After solving equations \eqref{vx} and \eqref{vy}, we obtain
\be
&&\label{Gammax}\Gamma_x=\frac{qE\tau_f\left(1-\omega_c^2\tau_f^2\right)}{1+\omega_c^2\tau_f^2}, \\ 
&&\label{Gammay}\Gamma_y=-\frac{2qE\omega_c\tau_f^2}{1+\omega_c^2\tau_f^2}
.\ee
Now, ansatz \eqref{ansatz} can be written as
\be\label{ansatz(1)}
f_f=f_f^0-qE\tau_f\frac{\partial f_f^0}{\partial p_x}-qE\tau_f\left(\frac{1-\omega_c^2\tau_f^2}{1+\omega_c^2\tau_f^2}\right)\frac{\partial f_f^0}{\partial p_x}+2qE\left(\frac{\omega_c\tau_f^2}{1+\omega_c^2\tau_f^2}\right)\frac{\partial f_f^0}{\partial p_y}
.\ee
By using $\frac{\partial f_f^0}{\partial p_x}=v_x\frac{\partial f_f^0}{\partial \omega_f}=-v_x\beta f_f^0\left(1-f_f^0\right)$ and $\frac{\partial f_f^0}{\partial p_y}=v_y\frac{\partial f_f^0}{\partial \omega_f}=-v_y\beta f_f^0\left(1-f_f^0\right)$, 
eq. \eqref{ansatz(1)} gets simplified into
\be\label{ansatz(2)}
\nonumber f_f &=& f_f^0+qE\tau_fv_x\beta f_f^0\left(1-f_f^0\right)+qE\tau_fv_x\beta\left(\frac{1-\omega_c^2\tau_f^2}{1+\omega_c^2\tau_f^2}\right)f_f^0\left(1-f_f^0\right) \\ &&-2qEv_y\beta\left(\frac{\omega_c\tau_f^2}{1+\omega_c^2\tau_f^2}\right)f_f^0\left(1-f_f^0\right)
.\ee
This leads to the determination of $\delta f_f$ as
\be\label{deltaf.(q)}
\delta f_f=2qEv_x\beta\left(\frac{\tau_f}{1+\omega_c^2\tau_f^2}\right)f_f^0\left(1-f_f^0\right)-2qEv_y\beta\left(\frac{\omega_c\tau_f^2}{1+\omega_c^2\tau_f^2}\right)f_f^0\left(1-f_f^0\right)
.\ee

\renewcommand{\theequation}{B.\arabic{equation}}
\section{Derivation of equation \eqref{deltaf.q1}}\label{I.C. of Q.D.F.2}
Substituting the value of $L$ \eqref{L} in eq. \eqref{eq.2} and simplifying, we have
\be
&& \nonumber\frac{\left(\omega_f-h_f\right)}{T}v_x\left(\partial^xT-\frac{T}{nh_f}\partial^xP\right)+\frac{\Gamma_xv_x}{\tau_f}-\omega_c\Gamma_yv_x-qEv_x \\ && \nonumber+\frac{\left(\omega_f-h_f\right)}{T}v_y\left(\partial^yT-\frac{T}{nh_f}\partial^yP\right)+\frac{\Gamma_yv_y}{\tau_f}+\omega_c\Gamma_xv_y+\tau_f\omega_cqEv_y \\ && +p_0\frac{DT}{T}-\frac{p^\mu p^\alpha}{p_0}\nabla_\mu u_\alpha+TD\left(\frac{\mu_f}{T}\right)=0
.\ee
Equating the coefficients of $v_x$ and $v_y$ on both sides of the 
above equation, we obtain 
\be
&&\label{eq.3}\frac{\left(\omega_f-h_f\right)}{T}\left(\partial^xT-\frac{T}{nh_f}\partial^xP\right)+\frac{\Gamma_x}{\tau_f}-\omega_c\Gamma_y-qE=0, \\ 
&&\label{eq.4}\frac{\left(\omega_f-h_f\right)}{T}\left(\partial^yT-\frac{T}{nh_f}\partial^yP\right)+\frac{\Gamma_y}{\tau_f}+\omega_c\Gamma_x+\tau_f\omega_cqE=0
.\ee
By solving equations \eqref{eq.3} and \eqref{eq.4}, $\Gamma_x$ and 
$\Gamma_y$ are respectively determined as
\be
\label{Gamma(x)}\nonumber\Gamma_x &=& \frac{qE\tau_f\left(1-\omega_c^2\tau_f^2\right)}{1+\omega_c^2\tau_f^2}-\frac{\tau_f\left(\omega_f-h_f\right)}{T\left(1+\omega_c^2\tau_f^2\right)}\left(\partial^xT-\frac{T}{nh_f}\partial^xP\right) \\ && -\frac{\omega_c\tau_f^2\left(\omega_f-h_f\right)}{T\left(1+\omega_c^2\tau_f^2\right)}\left(\partial^yT-\frac{T}{nh_f}\partial^yP\right), \\ 
\label{Gamma(y)}\nonumber\Gamma_y &=& -\frac{2\omega_c\tau_f^2qE}{1+\omega_c^2\tau_f^2}-\frac{\tau_f\left(\omega_f-h_f\right)}{T\left(1+\omega_c^2\tau_f^2\right)}\left(\partial^yT-\frac{T}{nh_f}\partial^yP\right) \\ && +\frac{\omega_c\tau_f^2\left(\omega_f-h_f\right)}{T\left(1+\omega_c^2\tau_f^2\right)}\left(\partial^xT-\frac{T}{nh_f}\partial^xP\right)
.\ee
Using the values of $\Gamma_x$ and $\Gamma_y$ in ansatz \eqref{ansatz} 
and then simplifying, we get the infinitesimal change of the quark 
distribution function as
\be\label{deltaf.(q1)}
\nonumber\delta f_f &=& \frac{2qE\tau_fv_x\beta f_f^0\left(1-f_f^0\right)}{1+\omega_c^2\tau_f^2}-\frac{2qE\omega_c\tau_f^2v_y\beta f_f^0\left(1-f_f^0\right)}{1+\omega_c^2\tau_f^2}-\beta^2 f_f^0\left(1-f_f^0\right)\frac{\tau_f(\omega_f-h_f)}{\left(1+\omega_c^2\tau_f^2\right)} \\ && \nonumber\times\left[v_x\left(\partial^xT-\frac{T}{nh_f}\partial^xP\right)+v_y\left(\partial^yT-\frac{T}{nh_f}\partial^yP\right)\right]-\beta^2 f_f^0\left(1-f_f^0\right) \\ && \times\frac{\omega_c\tau_f^2(\omega_f-h_f)}{\left(1+\omega_c^2\tau_f^2\right)}\left[v_x\left(\partial^yT-\frac{T}{nh_f}\partial^yP\right)-v_y\left(\partial^xT-\frac{T}{nh_f}\partial^xP\right)\right]
.\ee

\end{appendices}

\end{document}